\documentstyle[12pt]{article}
\begin{document}
\renewcommand{\thefootnote}{\fnsymbol{footnote}}
\def\thefootnote{\arabic{footnote}}
\setcounter{footnote}{1}
\def\theequation{\arabic{section}.\arabic{equation}}
\def\thesection{\Roman{section}.}

\begin{titlepage}
\parindent 0mm
\topskip 20mm

\hspace{10cm}  KLTE-DTP/1994/1

\vspace{15mm}

\begin{center}
\large
{\bf
Phase structure of $SU(2)$ Yang-Mills theory with global center
symmetry
}

\vspace{1cm}

\normalsize
\baselineskip 6mm
K. Sailer\footnotemark[1]

\vspace{6mm}

{\em Department for Theoretical Physics\\
Kossuth Lajos University\\
H-4010 Debrecen, Pf. 5, Hungary  }

\footnotetext[1]{
E-mail: sailer@tigris.klte.hu
                  }
\end{center}


\begin{abstract}
\normalsize
\rm
\parindent 10mm
\baselineskip  5mm

Retaining only the `timelike' component  $A_0$ of the
vector potential
a skelet model with explicit global center symmetry is constructed
 for  $SU(2)$ Yang-Mills theory.
It is shown that the $A_0$ gluon vacuum is equivalent with the
4-dimensional Coulomb-gas.
In 1--loop approximation,
 the effective theory of the skelet model exhibits non-local
self-interaction. A coupling constant $\lambda$ is found that
governs the loop expansion. For $\lambda < 1$
the effective theory does not confine (the string tension vanishes),
whereas for $\lambda = 1$ confinement (with non-vanishing string
 tension)
 takes place. In the deconfined phase the $A_0$ gluons exhibit
non-zero rest mass and global center symmetry is broken by the
vacuum
state.
 Confinement sets on for $\lambda =1$ when the rest
mass of $A_0$ gluons vanishes and the global center symmetry of the
vacuum is restored.

PACS number(s): 11.15.Tk, 11.10.Lm
\end{abstract}

\end{titlepage}

\newpage
\normalsize
\rm
\textheight 21.0cm
\headheight -1. cm
\oddsidemargin 1mm
\leftmargin 1mm
\topmargin -5mm
\textwidth 15.5cm
\parindent 10mm
 \baselineskip 6.0mm
\topskip 0mm
\footheight 1.5cm
\footskip 5mm

\section{INTRODUCTION}

Lattice QCD results indicate that confinement  and global center
symmetry are strongly correlated\cite{Wil74,Pol88,Pol91,Pol92}.
Simulating pure Yang-Mills systems on the lattice it was found:
{\it {(i)}} there exists a critical temperature $T_c $
for quarkless QCD at which phase transition takes place;
{\it {(ii)}} the free energy of a static $q {\bar q}$ pair rises
linearly with increasing separation distance and the string tension
obtained from the Wilson loop and that from the correlation of
Polyakov lines agree well for $T < 0.6 T_c$; {\it {(iii)}} the
confining phase maintains global center symmetry, but it is
broken in the deconfined phase. It is the most important
message of the lattice results for us that
the  transition from the confined
to the deconfined phase seems to be accompanied by
breaking of the global center
symmetry. In the present article
we  investigate the continuum $SU(2)$ gauge theory with global
center symmetry. We  show that the confined phase (with
 non-vanishing string
tension) may exist with  global center symmetry, whereas in the
deconfined phase (with vanishing string tension)  this symmetry
is broken.
In connection with the phase structure of the theory
we also show that only a few of the infinitely many parameters are
of significance.

Our treatment can be outlined as follows. We start with the
 generating
functional of the Green's functions of the lattice regularized
$SU(2)$ Yang-Mills theory.
In order to ensure  global center symmetry the $SU(2)$ invariant
Haar measure is used in the path
integral\cite{Joh91}, and the gauge fixing is performed
without violating global center symmetry. This leads to a cut-off
dependent periodic effective potential. For the sake of simplicity
 we
neglect the spatial components  of the vector potential remaining
invariant under global center transformations. The 'skelet' $SU(2)$
Yang-Mills theory obtained in this way contains only the timelike
$A_0$
components of the vector potential being sensitive to the global
center
transformations.

We show that the skelet model is equivalent with the macrocanonical
ensemble for a $D=4$ gas consisting of infinitely heavy
point charges interacting via Coulomb
 interaction.
This structure of the vacuum is completely induced by global center
symmetry.
 In addition to the strong coupling constant the model
contains infinitely many other parameters: the fugacities of the
charges
$\pm 1$, $\pm 2$, etc.

We expand the effective potential at its minima and derive the
 effective
action in 1-loop approximation\cite{Zin89}. The 1--loop
order effective action contains a non-local  self-interaction.
A parameter $\lambda$ is found governing the loop expansion.

In the framework of the effective theory
 we derive an expression for the string tension.
 The string
tension is determined through the propagator of the gauge field in
 the
effective theory. According to the value of the loop expansion
parameter $\lambda$
 the propagator of the effective
theory has qualitatively different behaviour. For $\lambda < 1$
it exhibits
simple poles whereas for $\lambda = 1$ a double pole occurs. In
the first case
the string tension vanishes, while it has a non-vanishing value
in the second case.

\section{SKELET MODEL}
\setcounter{equation}{0}

The generating functional $Z \lbrack j \rbrack$ of the $SU(2)$
Yang-Mills theory is by definition the vacuum to vacuum transition
amplitude in the presence of the external source $j_\mu^a$:
\begin{eqnarray}
  Z \lbrack j \rbrack \sim \int {\cal D} A_i^a \int {\cal D} A_0^a
  \exp \left\{ - \frac{1}{4g^2} \int d^4 x F_{\mu \nu}^a
   F_{\mu \nu}^a
        - S_{gf} - \int d^4 x j^a_\mu A_\mu^a - S_{gh}  \right\}
\end{eqnarray}
with the gauge fixing term $S_{gf}$ and the ghost term $S_{gh}$.
We modify the generating functional of the usual perturbative
treatment in order to ensure its invariance under global center
transformations following \cite{Pol91,Pol92,Joh91}, but we take
global center transformations into account in a rather explicit way
and use gauge conditions which do not violate global center symmetry.

\begin{enumerate}
\item

 Under an arbitrary finite $SU(2)$ gauge transformation $h(x)$
the spatial and the timelike components of the vector potential
transform as follows\cite{Pol92,Joh91}:
\begin{eqnarray}
\label{trafo}
  \left( A_i^a \frac{\tau^a}{2} \right)^h
  &=& h(x) \left( A_i^a \frac{\tau^a}{2} \right) h^{-1} (x)
    + h(x) \partial_i h^{-1} (x)  , \nonumber\\
  \left(  \exp \left\{ {\rm i} a A_0^a \frac{\tau^a}{2}  \right\}
  \right)^h
   &=&
  \left(  \exp \left\{ {\rm i} a A_0^a \frac{\tau^a}{2}  \right\}
  \right)  h(x)  .
\end{eqnarray}

The center of a group consists of the elements
commuting with any element of the group. The center of the group
 $SU(2)$
consists of the elements $z_0 =1$ and $z_1=-1$.  Global center
 transformation
means the same center transformation at every point of 3-space at a
given
time slice. This transformation takes advantage of the fact that we
 can
perform finite gauge transformations at any time slice separately.
In the continuum limit this leads to field configurations
discontinuous
in time. The global center transformation $z_1 =-1= \exp \left\{
{\rm i} 2\pi \frac{\tau^3}{2} \right\}$ is not trivial. Parametrizing
the general group element $h (\alpha {\vec n} ) = \exp \left\{
{\rm i} \alpha {\vec n} \frac{ \vec \tau}{2}  \right\}$ by the colour
vector  ${\vec \alpha} = \alpha {\vec n}$
($\alpha \in \lbrack 0, 2\pi \rbrack$, and ${\vec n}$ arbitrary
 unit vector
 in the solid angle $4\pi$), we get from Eq. (\ref{trafo}) for
 the effect
of the center transformation:
\begin{eqnarray}
\label{glocen}
   h \left( \alpha {\vec n} \right) z_1 &=&
   h \left( - (2\pi - \alpha ) {\vec n}  \right)  .
\end{eqnarray}
The colour vector $\alpha {\vec n}$ flips
to $ - (2\pi -\alpha ) {\vec n}$ due to the non-trivial
center transformation $z_1$.
In order to treat the field configurations obtained by non-trivial
global center transformation explicitly,
we  allow for
global center transformations $z_1^{k_t}$ at each time slice $t$
($k_t =0$ for identity transformation and $k_t =1$ for $z_1 =-1$).
Taking the parameter of the local gauge transformation from
 the sphere $S_{2\pi }$  of radius $2\pi$,
${\vec \alpha}_{{\vec x}t} \in S_{2\pi} $,
and taking the sum over $k_t =0, 1$ we cover the
 $SU(2)$ group twice.
The multiple counting of the gauge equivalent
intermediate vacuum states shall be
removed by dividing the path integral by the appropriate multiple of
the volume of the gauge group.

The physical vacuum to vacuum transition amplitude
 ${\cal Z} \lbrack 0 \rbrack$ is constructed
 in Hamiltonian formalism starting with the gauge
 $A_0^a \equiv 0$ $(a=1,2,3)$
 \cite{Pol92,Joh91} (see Appendix A).
 The whole interval $T$ of time evolution
 is divided into
small intervals of length $a$. At each time slice the projection
 to colour singlet vacuum states is performed by the operator:
\begin{eqnarray}
  P_\theta &=& \sum_{k_t =0, 1}
    \int {\cal D}_H (hz^{k_t} ) \; (hz^{k_t} )
          =   \sum_{k_t =0, 1}
 \int {\cal D}_H h \; (hz^{k_t} )
                   \nonumber\\
    &=&
   \sum_{k_t =0,1}
   \prod_{\vec x}  \int_{S_{2\pi} }  d_H {\vec \alpha}_{{\vec x}t}
   \;   h \left( ( \alpha_{{\vec x}t}
       + 2\pi k_t (-1)^{k_t} ) {\vec n}_{{\vec x}t}
 \right) ,
\end{eqnarray}
with the gauge invariant Haar measure:
\begin{eqnarray}
   d_H {\vec \alpha} &=& d^3 \alpha \frac{1}{4\pi^2}
       \frac{ \sin^2 \left( \frac{1}{2} \alpha \right)  }{ \alpha^2
                                                         }
\end{eqnarray}
($d^3 \alpha$ the flat measure).
Finally the path integral is divided by the multiple of the volume
of the gauge group:
\begin{eqnarray}
   \sum_{  \{ k_t =0,1 \} } \prod_{ {\vec x}t} \int_{ S_{2\pi }}
         d_H {\vec \alpha}_{ {\vec x} t}
                 & \equiv &  \int_{\cal G} {\cal D}_H h.
\end{eqnarray}
Later we make use of the integral
\begin{eqnarray}
 \int_{\cal G} {\cal D}_H  (aA^{0a} ) \Phi \lbrack aA^{0a} \rbrack
   & =&  \sum_{ \{ k_t =0,1 \} } \prod_{ {\vec x}t}
    \int_{S_{2\pi } }
    d_H (aA^{0a} )_{ {\vec x}t} \Phi \lbrack aA^{0a}_{ {\vec x}t }
     + 2\pi k_t \rbrack ,
\end{eqnarray}
with arbitrary functional $\Phi \lbrack aA^{0a } \rbrack$.
We can reintroduce the zeroth component of the vector
 potential\cite{Pol92}
 via
\begin{eqnarray}
\label{A0}
 A_{0 \; {\vec x} t}^a = g \alpha_{{\vec x}t}^a /T .
\end{eqnarray}

\item
We introduce the gauge fixing conditions $F^a \lbrack A_0^{h \; a} ,
  A_j^{h \; a} \rbrack$ in a gauge invariant way by inserting
\begin{eqnarray}
\label{unity}
  1 &=& \Delta \lbrack A \rbrack \int_{\cal G} {\cal D} h
    \prod_{a=1}^3 \delta \lbrack F^a \lbrack A^{h} \rbrack
                         \rbrack
\end{eqnarray}
in the path integral. The functional $\Delta \lbrack A \rbrack$
is invariant under gauge transformations including global center
transformations as well. Making use of the gauge (and center)
 invariance
of the integration measure, and the $\Delta$ functional, we can
factorize out the multiple of the group volume $\int_{\cal G}
{\cal D} h$ and can divide with it. Thus we find for the physical
vacuum to vacuum transition amplitude Eq. (\ref{vacvac}):
\begin{eqnarray}
  {\cal Z} \lbrack 0 \rbrack  \sim
  \int_{\cal G} {\cal D} (aA_0^a )    \int {\cal D} A_i^a
  \left. \Delta \lbrack A \rbrack \right|_{ F \equiv 0}
  \left( \prod_{a=1}^3 \delta \lbrack F^a \lbrack A \rbrack \rbrack
  \right)
  e^{ {\rm i} S \lbrack A^{h (A_0 ) \; j } , A^j \rbrack   } ,
\end{eqnarray}
with the action\cite{Pol92}:
\begin{eqnarray}
   S \lbrack A^{h (A_0 ) \; j } , A^j \rbrack    &=&
  - \frac{1}{4g^2} \int_{-T/2}^{T/2}
   \int d^3 x {\bar F}_{\mu \nu} {\bar F}^{\mu \nu} ,
\end{eqnarray}
and with the generalized field strength tensor:
\begin{eqnarray}
  {\bar F}^{j0 \; a}_{{\vec x}t} = - \frac{1}{a}
  \left( A_{ {\vec x} \; t+a }^{h \; ja} - A_{ {\vec x} t}^{ja}
  \right)    ,
  \qquad  {\bar F}^{ij \; a} = F^{ij \; a}  .
\end{eqnarray}
For infinitesimal gauge transformations the generalized field
strength tensor takes the usual form.

\item
We complete our definition of the vacuum to vacuum transition
 amplitude
by a particular center invariant choice of the gauge conditions.
According to (\ref{glocen}) the colour vector $A_0^a$ does not
change
its orientation in colour space due to center transformation.
Further on colour vectors $aA_0 {\vec n} $
and $ ( aA_0 - 2\pi  ) {\vec n}$
are center equivalent.
Therefore the gauge conditions
\begin{eqnarray}
  F^1 \lbrack A \rbrack = a A^1_{0 \; {\vec x}t} \equiv 0 , \qquad
  F^2 \lbrack A \rbrack = a A^2_{0 \; {\vec x}t} \equiv 0,
              \nonumber\\
  F^3 \lbrack A \rbrack = {\cal F} ( aA_0^3 )_{ {\vec x} \; t+a} -
                          {\cal F} ( aA_0^3 )_{ {\vec x} \; t}
          \equiv 0
\end{eqnarray}
do not violate global center symmetry for ${\cal F} (\alpha )$
periodic, ${\cal F} (\alpha ) = {\cal F} (\alpha + 2\pi )$.
 The gauge condition $F^3 =0$ is
the discretized version of the condition $a \partial_0 {\cal F}
 (aA_0 ) =0
$, and can be considered as the center invariant generalization of
the temporal gauge $\partial_0 A_0 =0$. This is a generalization
allowing for finite jumps of the zeroth component of the vector
potential at any of the time slices due to global center
transformation. As for small fields we would like to recover the
temporal gauge, we require ${\cal F} (\alpha ) \sim \alpha $ for
$\alpha \to 0$. The gauge condition $F^3 =0$ exhibits as a rule
many solutions, many Gribov copies exist. To avoid this complication
we require that the Gribov copies are those field configurations
obtained by global center transformation. It is achieved if the
equation ${\cal F} (\beta ) - {\cal F} (\alpha ) =0$ has only the
single solution $\beta = \alpha$ for $\alpha , \; \beta \in \lbrack
0, 2\pi \rbrack$. For example the function
${\cal F} (\alpha ) = \tan  \left( \frac{1}{2} \alpha \right)$
satisfies all these requirements. In the derivation of the vacuum to
vacuum transition amplitude, however,  we have to make use of the
properties of the function ${\cal F} (\alpha )$, but we do not have
to specify it.
 The sums over the global center equivalent configurations (over
$k_t$'s) are just those over the Gribov copies in the path integral.

\end{enumerate}

The $\Delta \lbrack A \rbrack$ functional
defined above takes the following explicit form
(see Appendix B):
\begin{eqnarray}
\label{Delta}
 \left.  \Delta \lbrack A \rbrack  \right|_{F \equiv 0} &=&
  \exp \left\{ TV \ln (4\pi^2 ) -
   \sum_{ {\vec x}t} \frac{1}{2} \ln \sin^2
        \left( \frac{1}{2} a A_{0 \; {\vec x}t} \right)
   + \sum_{ {\vec x} t} \ln
        \left| {\cal F} ' ( a A_{0 \; {\vec x}t}^3 ) \right|
      \right.
                     \nonumber\\
     & &  \left.
     - \sum_{ {\vec x}t} \ln \left\lbrack
      \frac{1}{ ( aA_{0 \; {\vec x}t} )^2 } +
      \frac{1}{ ( 2\pi - aA_{0 \; {\vec x}t} )^2 } \right\rbrack
       \right\} ,
\end{eqnarray}
$ {\cal F}' (\alpha ) = \frac{d}{d\alpha } {\cal F} (\alpha )$.
Inserting this into the path integral ${\cal Z} \lbrack 0 \rbrack$,
we carry out the integration over $A_0^1$ and $A_0^2$ and also the
sums over $k_t$'s (see Appendix C).
Making use of the global center invariance of the   function ${\cal
F} (\alpha )$ and that of the action,
 we can  replace $\delta ( F^3 )$ by (see Eq. (\ref{FtoG}) ):
\begin{eqnarray}
  \frac{2 \sin^2 \left( \frac{1}{2} a A_{0 \; {\vec x}t}^3 \right)
}{ \left| {\cal F}' (a A_{0 \; {\vec x}t}^3 ) \right| }
 \delta \left(
        {\cal G}_{ {\vec x} \; t+a}
     -  {\cal G}_{ {\vec x} \; t}  \right)
\end{eqnarray}
with ${\cal G} ( \alpha ) = \frac{1}{2} ( \alpha - \sin \alpha )$.
This allows us to recover the Haar measure and eliminate the
dependence on the function ${\cal F} (\alpha )$.
Shifting the argument of the Dirac delta
 by a field independent value $c_{ {\vec x}t}$ and integrating
over it by Gaussian weights, we obtain (Appendix C):
\begin{eqnarray}
 {\cal Z} \lbrack 0 \rbrack &=&
 \int {\cal D} u \int {\cal D} A_i^a
  \exp \left\{ {\rm i} S \lbrack A^{ h(u) \; j} , A^j \rbrack
          + \sum_{ {\vec x}t} \ln 2
      +  \sum_{ {\vec x}t} \ln \sin^2 \left( \frac{1}{2} au \right)
                \right.
                   \nonumber\\
& &    \left.
   - \frac{1}{2g^2 \xi } \sum_{ {\vec x} t} \left(
    \frac{1}{2} (  au -
    \sin (au ) \; )_{ {\vec x} \; t+a } -
    \frac{1}{2} (  au -
    \sin (au ) \; )_{ {\vec x} t}         \right)^2
       \right\}  ,
\end{eqnarray}
where $au \equiv aA_0^3 \in \lbrack -2\pi , 2\pi \rbrack$.

We see that retaining global center symmetry has lead to the
periodic
tree level effective potential due to the Haar measure, as
expected\cite{Pol88,Pol91,Pol92}.

We see
from Eq. (\ref{trafo})
  that the spatial components of the vector potential remain
unaltered under global center transformations. For our goal is to
investigate the consequences of the global center symmetry we
 neglect
the spatial components of the vector potential by
setting $A_i^a =0$ in the present work.

In this way we obtained
the `skelet' model which only contains those  degrees of freedom
being sensitive to the global center transformations. In Euclidean
space the
generating functional  of the Green's functions
  takes the form:
\begin{eqnarray}
  Z \lbrack q \rbrack &=& {\cal Z} \lbrack q \rbrack / {\cal Z}
    \lbrack 0 \rbrack  ,
\end{eqnarray}
with
\begin{eqnarray}
  {\cal Z} \lbrack q \rbrack &=& \int {\cal D} u
   e^{ - S_s - S_{gf} - \int d^4 x q u } ,
\end{eqnarray}
and $ a^3 j_0^3 \equiv q$ the  external source coupled to $u$,
where
\begin{eqnarray}
  S_s &=& \frac{1}{2g^2} \int d^4 x ( \partial^i u )^2
       + {\cal V} \lbrack u \rbrack ,
           \\
   S_{gf} &=& \frac{1}{2g^2 \xi } \int d^4 x \frac{1}{a^2}
   \left( \partial_0 {\cal G} (au) \right)^2  ,
\end{eqnarray}
and the tree level effective potential is given by:
\begin{eqnarray}
  {\cal V} = \int d^4 x V (u (x)) ,
             \qquad
   V (u) = - \frac{1}{a^4} \ln \sin^2 \left( \frac{1}{2} au \right)
             - \ln 2 .
\end{eqnarray}

As the action $S_s + S_{gf}$ exhibits $u \to -u$ symmetry, we can
rewrite the integrals:
\begin{eqnarray}
   \prod_x \int_{ -2\pi /a}^{2\pi /a} du_x \ldots
   \to \prod_x 2 \int_0^{ 2\pi /a} du_x \ldots
   = e^{ TVa^{-4} \ln 2 }  \prod_x \int_0^{2\pi /a} du_x  \ldots
\end{eqnarray}
The additional exponential factor just cancels the zero frequency
term of the Fourier expansion of the tree level effective potential.

As the tree level effective potential has minima at $au =\pm \pi$,
the field configurations close to these contribute most
 significantly
to the path integral. Therefore we use the expansion
\begin{eqnarray}
   {\cal G} ( \alpha ) & = & {\cal G} ( \pm \pi ) +
       \left.  \frac{ d {\cal G} }{ d \alpha }  \right|_{ \alpha
       = \pm        \pi }  ( \alpha \mp \pi )
        + {\cal O} ( (\alpha \mp \pi )^3
                     )
                \nonumber\\
      &=&  \alpha \mp \frac{1}{2} \pi
       + {\cal O} ( ( \alpha \mp \pi )^3
                    ) ,
\end{eqnarray}
which
choosing $\xi =1$ leads to:
\begin{eqnarray}
   S_{gf} &=& \frac{1}{2g^2} \int d^4 x ( \partial_0 u )^2  .
\end{eqnarray}
Then we obtain for the generating functional:
\begin{eqnarray}
  {\cal Z} \lbrack q \rbrack &=& \int {\cal D} u
   e^{ - S \lbrack u \rbrack  - \int d^4 x q u } ,
\end{eqnarray}
with the action
\begin{eqnarray}
 S \lbrack u \rbrack &=&
  \frac{1}{2g^2} \int d^4 x \left\lbrack
  ( \partial^i u )^2 + ( \partial_0 u )^2  \right\rbrack
  + {\cal V}_0 \lbrack u  \rbrack
\end{eqnarray}
and with the tree level effective potential
\begin{eqnarray}
  {\cal V}_0 &=& {\cal V} - TV a^{-4}  \ln 2 = \int d^4 x V_0 (u).
\end{eqnarray}
The Fourier expansion of the periodic effective potential is given
by:
\begin{eqnarray}
\label{Four}
  V_0 (u) &=& \frac{1}{a^4} \sum_{ \sigma \neq 0} v_\sigma
               \cos (  \sigma a u) ,
\end{eqnarray}
with $v_\sigma = v_{-\sigma} =   |\sigma |^{-1} $.
The tree level effective potential is periodic in the interval
$-2\pi \le au \le 2\pi$ with minima at $au =\pm \pi$, and it
tends to $+\infty$ for $au=0$,  $ \pm 2\pi$.

\section{ANALOGY WITH THE COULOMB GAS}
\setcounter{equation}{0}

Having performed the Fourier expansion of the effective potential
and the Taylor expansion of $\exp \{ - {\cal V}_0 \}$, we carry out
the Gaussian path integral over the field $u$ and
obtain\cite{Joh91}:
\begin{eqnarray}
  {\cal Z}_C \lbrack q \rbrack &=&
   \sum_{n=0}^\infty  \frac{ (-1)^n}{n!}
    \left( \prod_{j=1}^n \sum_{x_j} \sum_{ \sigma (j) \neq 0}
            v_{\sigma (j) }  \right)
            z_C \lbrack q \rbrack ,
\end{eqnarray}
with
\begin{eqnarray}
    z_C \lbrack q \rbrack &=&  \left( {\mbox {Det}} D
     \right)^{-1/2}
    \exp \left\{  \frac{1}{2}  \sum_{xy}
      \left(
       \sum_{j=1}^n   {\rm i} \sigma (j) \delta_{ x x_j} -q_x
       \right)   a^2 D_{xy}   \cdot
                \right.    \nonumber\\
     & &  \cdot \left.
      \left(
       \sum_{k=1}^n   {\rm i} \sigma (k) \delta_{ y x_k} -q_y
       \right)
          \right\}  .
\end{eqnarray}
For vanishing external source
\begin{eqnarray}
  {\cal Z}_C \lbrack 0 \rbrack &=&  ( {\mbox {Det}} )^{- 1/2}
   \sum_{n=0}^\infty  \frac{ (-1)^n}{n!}
    \left( \prod_{j=1}^n \sum_{x_j} \sum_{ \sigma (j) \neq 0}
            v_{\sigma (j) }  \right)
     \exp \left\{ - \frac{1}{2} \sum_{ x_j x_k } \sigma (j)
           D_{ x_j x_k }  \sigma (k)   \right\}
            \nonumber\\
\end{eqnarray}
is identical with the macrocanonical partition sum
 of the $D=4$ dimensional Coulomb gas
with infinitely heavy (static) point charges $\sigma =\pm 1$,
 $\pm 2$,
$\ldots$.  The Fourier coefficients
$-v_{ \sigma }$  of the periodic
tree level effective potential
play the role of the fugacities of the charges $\sigma  $.
The free $u$-propagator,
\begin{eqnarray}
    a^2 D (x_j , x_k ) & \to & \frac{ a^2 g^2 }{
     \pi^2 ( x_j - x_k )^2        }
  \qquad  {\mbox {for}} \qquad a \to 0
\end{eqnarray}
represents the Coulomb interaction between the static charges.
In addition to the strong coupling constant $g$ the model contains
infinitely many other parameters, the fugacities $v_\sigma$.

We conclude that the requirement of global center symmetry leads
 to the
equivalence of the $A_0$ gluon vacuum with a $D=4$ dimensional
Coulomb gas of static charges.

\section{EFFECTIVE THEORY}
\setcounter{equation}{0}

For now on we use lattice
regularization which is rather natural from the point of view
of the
definition of the path integral and of the dependence of the
effective potential on the lattice size $a$. Although the
 calculations
have been done in lattice regularized form,
 for the sake of simplicity we write the expressions
in their continuous form if it is not misleading.
The main steps of obtaining the effective
theory are as follows:

\begin{enumerate}

\item
 The effective potential at the tree level has
minima
at $a  u =\pm \pi$. Therefore
 the vacuum expectation value $\langle u(x) \rangle
\equiv {\bar u}$ does not vanish as a rule, but corresponds to
one of
the minima of the effective potential.  Let us shift the field
 variable
\begin{eqnarray}
    u(x) &=& {\bar u} + \eta (x) ,
\end{eqnarray}
and write the generating functional of the Green's functions as:
\begin{eqnarray}
  {\cal Z} \lbrack q \rbrack &=&
   e^{ a {\bar u} a^{-4} \int d^4 x \; q }
  \int {\cal D} \eta
 \exp \left\{ \frac{1}{2g^2} \int d^4 x \eta \Box
   \eta            \right.
                \nonumber\\
    & &  \left.
- a^{-4} \int d^4 x \eta q - \int d^4 x V_0 ( {\bar u} +\eta )
                          \right\} .
\end{eqnarray}

\item
Now we can linearize the exponential of the path integral
in terms of $\eta (x)$
 by introducing
the auxiliary field $h(x)$ by the definition:
\begin{eqnarray}
\lefteqn{
   \exp \left\{ \frac{1}{2g^2 } \int d^4 x \eta \Box \eta \right\}
   =   \left( {\mbox {Det}} D \right)^{1/2}   \cdot
        }                \nonumber\\
  & & \cdot
\int {\cal D} h \exp \left\{
     \frac{1}{2a^8} \int d^4 x d^4 y h(x)  a^2 D (x,y) h(y)
    + \int \frac{d^4 x }{a^4} h(x) a \eta (x) \right\} .
            \nonumber\\
\end{eqnarray}
The generating functional has now the form:
\begin{eqnarray}
  {\cal Z} \lbrack q \rbrack &=&
  \left( {\mbox {Det}} D \right)^{1/2}
   e^{ a {\bar u} a^{-4} \int d^4 x q }
  \int {\cal D} h
   e^{ \frac{1}{2a^8} \int d^4 x d^4 y h(x) a^2 D(x,y) h(y) }
                \cdot      \nonumber\\
       & &   \cdot
  \int {\cal D}     \eta
  e^{ a^{-4} \int d^4 x \eta (h-q) - \int d^4 x V ( {\bar u} +\eta )
       }   .
\end{eqnarray}

In order to clarify the meaning of the auxiliary field
let us go back to
Minkowski space, which  roughly means the replacements
 $d^4 x \to -{\rm
i} d^4 x$, $\eta \to {\rm i} \eta$, $q \to {\rm i} q$,
 $h \to {\rm i} h$, and forget about
lattice regularization. Then the naive integration over
${\cal D} \eta$ leads to the Dirac-delta  functional:
\begin{eqnarray}
   \delta \left\lbrack h(x) - q (x) + \sum_j  \sigma (j)
    a \delta (x-x_j )
          \right\rbrack .
\end{eqnarray}
According to that the auxiliary field $h(x)$ can be interpreted
as the
effective charge density being the sum of the external charge
 density
and the polarisation charge density. The latter arises due to the
Coulomb charges in the vacuum.

\item
We expand the tree level effective potential in the Fourier series
(\ref{Four})  and the exponential $e^{ - {\cal V}}$ in Taylor series,
 and perform the path integral over $\eta$. In order to
integrate over all group elements in terms of the new variable
$\eta_x$, we choose the interval of integration as
$a\eta_x \in \lbrack -\pi , \pi  \rbrack$. Then we obtain:
\begin{eqnarray}
   {\cal Z} \lbrack q \rbrack &=&
   e^{ a {\bar u} a^{-4} \int d^4 x \; q }
  \int {\cal D} h
   e^{ \frac{1}{2a^8} \int d^4 x d^4 y h(x) a^2 D(x,y) h(y)
   + \ln z \lbrack h - q \rbrack   }  ,
\end{eqnarray}
where
\begin{eqnarray}
 e^{ \ln z \lbrack h-q \rbrack }  &=&
  \left( {\mbox {Det}} D \right)^{1/2}
  \sum_{n=0}^\infty \frac{ (-1)^n}{n!} \left( \prod_{j=1}^n \int
\frac{ d^4 x_j }{a^4} \sum_{\sigma (j) \neq 0} v_{ \sigma (j) }
            \right)
  I_\sigma^{(n)} \lbrack h-q \rbrack ,
                \nonumber\\
\end{eqnarray}
with
\begin{eqnarray}
  I_\sigma^{(n)} \lbrack \xi \rbrack &=&
  e^{ {\rm i}  \sum_{j=1}^n \sigma (j) a {\bar u}  }
  \prod_x \int_{-\pi }^{\pi } d (a\eta_x )
   e^{ (\xi_x + {\rm i} \sum_{j=1}^n \sigma (j) \delta_{xx_j }
       )  a\eta_x }
             \nonumber\\
  &=&
  \prod_x   2 \frac{ \sinh \left( \pi \xi_x \right)
                }{\xi_x}
  \cdot \prod_{j=1}^n
  e^{ {\rm i}   \sigma (j) ( a {\bar u} +\pi ) }
   \frac{  \xi_{x_j}  }{\xi_{x_j} + {\rm i} \sigma (j) } ,
\end{eqnarray}
and $ \xi_x =  h_x - q_x$. Here we used the correct discretized
expressions of the lattice regularized theory.
We can write:
\begin{eqnarray}
 \ln z \lbrack \xi \rbrack &=& \ln I_0 \lbrack \xi \rbrack -
   \ln z_0 \lbrack \xi \rbrack ,
\end{eqnarray}
with
\begin{eqnarray}
   \ln I_0 \lbrack \xi \rbrack &=&
   \sum_x \ln {\mbox {Det}} (a^2 D ) + \sum_x \left\lbrack
    \ln \sinh \left( \pi \xi_x \right)   -
    \ln \xi_x \right\rbrack  .
\end{eqnarray}
  The functional $z_0
\lbrack \xi \rbrack$
has the form of the macrocanonical partition sum of an ideal gas:
\begin{eqnarray}
 - \ln z_0 &=& \ln \left\{  \sum_{n=0}^\infty
  \frac{ (-1)^n }{ n!} \left( \sum_x \sum_{\sigma \neq 0}
   v_\sigma \frac{ \xi_x}{ \xi_x +  {\rm i} \sigma }
  e^{  {\rm i} \sigma (a {\bar u} + \pi ) }
                       \right)^n
                    \right\}
             \nonumber\\
  &=& - \sum_x \sum_{\sigma \neq 0} v_\sigma
  \frac{ \xi_x }{ \xi_x +  {\rm i} \sigma }
  e^{  {\rm i} \sigma ( a {\bar u} + \pi ) }   .
\end{eqnarray}
According to that we can interpret the Coulomb gas of charges
$\sigma$ as an ideal gas of quasiparticles in the external field
$h_x$.

\item To the next we perform the path integral over $h(x)$ by
the saddle
point method taking into account the tree level and the 1--loop
contributions. The solution of the saddle point equation
\begin{eqnarray}
\label{saddle}
   \frac{ \delta }{\delta h_x} \left\{ \frac{1}{2}
   \sum_{xy} h_x a^2 D_{xy} h_y +
    \ln z \lbrack h - q \rbrack \right\} =0
\end{eqnarray}
is a local functional $h_{0x} \lbrack q \rbrack$.
Performing the saddle point integration over the auxiliary field
$h_x$, the generating functional can be written as:
\begin{eqnarray}
   {\cal Z} \lbrack q \rbrack &=&
 e^{ - a {\bar u} \sum_x q_x }
   {\cal Z}_0 \lbrack q \rbrack {\cal Z}_1 \lbrack q \rbrack
    ,
\end{eqnarray}
where ${\cal Z}_0 \lbrack q \rbrack$ and ${\cal Z}_1 \lbrack q
\rbrack$ are the tree level and 1-loop contributions, respectively:
\begin{eqnarray}
 {\cal Z}_0 \lbrack q \rbrack &=&
    \exp \left\{ \frac{1}{2} \sum_{xy} h_{0x} \lbrack q \rbrack
     a^2 D_{xy} h_{0y} \lbrack q \rbrack
    +  \ln z \lbrack h_0 \lbrack q \rbrack -q \rbrack
    \right\} ,
\end{eqnarray}
\begin{eqnarray}
  {\cal Z}_1 \lbrack q \rbrack &=& \left\lbrack {\mbox {Det}}
   (a^2 Q )        \right\rbrack^{-1/2} ,
\end{eqnarray}
with
\begin{eqnarray}
  a^2 Q_{xy} &=& a^2 D_{xy}  + \frac{ \delta^2 \ln z \lbrack h_0 - q
\rbrack     }{  \delta h_{0x} \delta h_{0y} } .
\end{eqnarray}

\item
The generating functional $W \lbrack q
\rbrack = \ln Z \lbrack q \rbrack$ of the connected Green's functions
takes the following form:
\begin{eqnarray}
  W \lbrack q \rbrack &=& {\mbox {const.}} + W_0 \lbrack q \rbrack +
  W_1 \lbrack q \rbrack ,
\end{eqnarray}
where $W_0 \lbrack q \rbrack$, and $W_1 \lbrack q \rbrack $
are the tree level and 1-loop contributions, respectively,
for the Coulomb gas,
\begin{eqnarray}
  W_0 \lbrack q \rbrack &=& - a {\bar u} \sum_x q_x
   + \frac{1}{2} \sum_{xy} h_{0x } \lbrack q \rbrack
    a^2 D_{xy} h_{0y} \lbrack q \rbrack
    + \ln z \lbrack h_0 \lbrack q \rbrack - q \rbrack  ,
                \\
   W_1 \lbrack q \rbrack &=& - \frac{1}{2} {\mbox {Tr}} \ln (a^2 Q )
{}.
\end{eqnarray}

\item Let us derive the effective action in 1--loop order
defined as the Legendre transform of the generating functional
$W \lbrack q \rbrack$:
\begin{eqnarray}
\label{Gam}
  \Gamma \lbrack \chi \rbrack &=& - \sum_x a \left( \chi_x
  + {\bar u}    \right) q_x - W \lbrack q \rbrack ,
\end{eqnarray}
where the classical field $\chi_x$
  is given by
\begin{eqnarray}
\label{clas}
 a \chi_x + a {\bar u} &=& - \frac{ \delta W \lbrack q \rbrack
  }{ \delta   q_x }
\end{eqnarray}
 as the functional of the external source $q_x$.

\item
In order to make our treatment self-consistent we require
\begin{eqnarray}
\label{selfc}
   \langle a u_x \rangle =  a {\bar u} ,
  \qquad {\mbox {i.e.}}  \qquad
  \left.  a \chi_x   \right|_{ q=0 }  = 0.
\end{eqnarray}
  Furthermore we require that the effective potential
has minimum for the constant field configuration $u_x = {\bar u}$.
In this way we ensure that the action is expanded at the classical
solutions.

The solution of the saddle point equation (\ref{saddle})  and the
${\bar u}$ values satisfying the self-consistency equation
(\ref{selfc}) were found (Appendix D).  It was established
in 1-loop order that  the effective potential has extrema at
$au =0, \pm \pi , \pm 2\pi$. $a {\bar u} = \pm \pi$ correspond to the
minima of the tree level effective potential. Therefore we perform
 the expansion of the theory  around these configurations and
later check that the 1-loop effective potential has minima for
them. It is a consequence of global center symmetry and that of the
charge conjugation symmetry $u \to -u$, that the minima of the
effective potential remain at the same place if quantum fluctuations
are included.

The solution of the saddle point equation (\ref{saddle}) is given by
\begin{eqnarray}
\label{Solsp}
   h_{0x} - q_x &=& - (Q_0^{-1} D q )_x +
     \frac{1}{6} f_4 \left( Q_0^{-1} (Q_0^{-1} Dq)^3  \right)_x
    + {\cal O} (q^5 )  ,
\end{eqnarray}
with
\begin{eqnarray}
   (a^2 Q_0 )_{xy} &=& a^2 D_{xy} + f_2 \delta_{xy}  .
\end{eqnarray}

\item
We construct the effective action $\Gamma \lbrack \chi \rbrack$
by
inverting the functional dependence given by Eq. (\ref{clas}) and
inserting it in the r.h.s. of Eq. (\ref{Gam}) retaining terms up
to the fourth order in the classical field.

Although the skelet model contains infinitely many parameters,
the fugacities $v_\sigma$ and the strong coupling constant $g$,
the effective action in 1-loop order can be expressed in
terms of the parameters
\begin{eqnarray}
   f_2 =  2 \cdot 2! \sum_{\sigma > 0 }  v_\sigma ( \sigma )^{-2}
            > 0 ,   \qquad
   f_4 = -  2 \cdot 4! \sum_{\sigma > 0} v_\sigma ( \sigma )^{-4}
            < 0 ,
\end{eqnarray}
and the new coupling constant
\begin{eqnarray}
    \lambda &=& - \frac{f_4}{f_2 g^2}  >0 .
\end{eqnarray}
This is a significant reduction of the number of relevant parameters
of the theory.

Calculating the 1-loop contribution $W_1 \lbrack q
\rbrack$ to the generating functional of the connected Green's
functions in Appendix E, we show that it is of order $\lambda$.
Therefore we investigate the skelet model for $0 < \lambda \le 1$,
where the loop expansion seems to be justified.
The 1--loop contribution turned out to be convergent only if
$s \equiv 2f_2 /g^2 < 1/8$, so $s$ is a small parameter in the model.
 The expressions listed below are obtained by simple
algebraic manipulations from the tree and 1--loop contributions
to the generating functional, $W_0 \lbrack q \rbrack$ and
$W_1 \lbrack q \rbrack$, resp., given in Appendix E.

 The classical field Eq. (\ref{clas}) as the functional of
the external source is given by:
\begin{eqnarray}
   \chi_x &=& - f_2 \left( ( 1+  \lambda - f_2 D^{-1}
      + {\cal O} (\lambda^2 , s^2)  ) q \right)_x
   \nonumber\\
    & & - \frac{1}{6} f_4 q_x^3 + \frac{1}{6} f_4 f_2 ( D^{-1} q^3
)_x  + \frac{3}{4} f_4 f_2 q_x^2 (D^{-1} q)_x
    + {\cal O} ( \lambda^2 , q^5 ) .
\end{eqnarray}
Here $f_2 D^{-1}$ is of the order $s$.
For we did not go beyond the 1--loop approximation, the operator
in the linear term on the r.h.s. is known only up to ${\cal O}
(\lambda )$ terms. Therefore we must take its inverse with the same
accuracy. Then we get for  the inverse relation
\begin{eqnarray}
  q_x &=& - \left( \left( \frac{1-  \lambda }{f_2 } + D^{-1}
   \right) \chi        \right)_x
             \nonumber\\
       & & +
     \frac{ f_4}{6f_2^4} (1- 4  \lambda ) \chi_x^3
   - \frac{ f_4}{4f_2^3}  \chi_x^2 (D^{-1} \chi )_x
      + {\cal O} (\lambda^2 , \chi^5 ) ,
\end{eqnarray}
which was found by iteration first solving the equation without the
self-interaction terms.
The effective action takes now the form:
\begin{eqnarray}
  \Gamma \lbrack \chi \rbrack = \frac{1}{2} \sum_{xy} \chi_x
    \left( \frac{1 - \lambda  }{f_2}  \delta_{xy}
  + D^{-1}_{xy}  \right) \chi_y
   - \frac{f_4  (1-\lambda )^4}{24 f_2^4} \sum_x \chi_x^4  .
\end{eqnarray}
Here we included only the leading order local terms of the
self-interaction,
 taking into account that $ f_2 D^{-1} \sim
{\cal O} (s)$.

We see that $f_2 >0$, and $f_4 <0$ are the correct signs for the
effective potential having  minimum at $\chi_x \equiv 0$. That
justifies the expansion made.

The effective action describes a non-local theory with
quartic self-interaction.
We read off the inverse propagator of the $A_0^3$ gluons from the
quadratic term of the effective action:
\begin{eqnarray}
   {\tilde {\cal D}}_0^{-1} (p) &=&
   \frac{1- \lambda }{f_2} + {\tilde D}^{-1} (p)  + {\cal O}
(\lambda^2 , s^2 ) ,
\end{eqnarray}
with the lattice regularized inverse propagator
of the bare theory\cite{Par??}:
\begin{eqnarray}
 f_2 {\tilde D}^{-1} (p) &=& \frac{s}{ a^2} \sum_{ \mu =0}^3
  \left( 1   - \cos (ap^\mu )  \right)  .
\end{eqnarray}
 A rest mass is generated  due to the
periodic effective potential, i.e.  the global center symmetry
 of the
effective action.  This rest mass can be
read off by comparing the $p_\mu  \to  0$ limits of the inverse
propagator with the general form $g^{-2} ( p^2 + m^2 )$:
\begin{eqnarray}
\label{mass1}
 m^2 &=& a^{-2} \frac{2}{s} (1- \lambda ) =
    a^{-2} \frac{ -f_4}{f_2^2 } \frac{1-  \lambda }{\lambda }
      \ge 0 .
\end{eqnarray}
In the neighbourhood each of its minima
the effective potential at 1--loop order has the following form
(neglecting the off-diagonal non-local terms):
\begin{eqnarray}
\label{epot1}
  {\cal V}_1 \lbrack \chi \rbrack &=& \frac{1}{2g^2} m^2 a^2 \sum_x
      ( a\chi_x )^2
   - \frac{f_4  (1 - \lambda )^4}{4! f_2^4}
      \sum_x ( a \chi_x )^4 .
\end{eqnarray}
It is interesting to note that  the rest mass
as well as the quartic self-interaction vanish for $\lambda =1$ in
1-loop approximation. We show in the next chapter that the
 propagator
exhibits rather different analytic properties for
 $\lambda \neq 1$ and
$\lambda =1$, which lead to a non-vanishing string tension in
 the latter
case.

\end{enumerate}

\section{STRING TENSION}
\setcounter{equation}{0}

Let us determine the free energy of  a static quark-antiquark pair
positioned on the $z$ axis (the quark at $z=0$, the antiquark at
$z=L$). In the $SU(2)$ Yang-Mills theory the Wilson operator
is given by:
\begin{eqnarray}
  w ( L ; \lbrack A \rbrack ) &=&
   {\mbox {Tr}} \left(
   \left. e^{ - {\rm i} \int_T^0 dx^0 \frac{1}{2} gA_0^a \tau^a  }
   \right|_{ z=0}
   \left. e^{ - {\rm i} \int_L^0 dz \frac{1}{2} gA_z^a \tau^a  }
   \right|_{ x^0 =T}
             \right.   \cdot
             \nonumber\\
    & & \cdot    \left.
   \left. e^{ - {\rm i} \int_0^T dx^0 \frac{1}{2} gA_0^a \tau^a  }
   \right|_{ z=L}
   \left. e^{ - {\rm i} \int_0^L dz \frac{1}{2} gA_z^a \tau^a  }
   \right|_{ x^0 =0}
                  \right)
\end{eqnarray}
which leads in the skelet model to the expression:
\begin{eqnarray}
 w ( L ; \lbrack u \rbrack ) &=& {\mbox {Tr}}
  \exp  \left\{  {  \rm i} \int_0^T  dx^0  \frac{1}{2} \left(
   u (x^0 , {\vec 0} , 0 ) - u ( x^0 , {\vec 0} , L)  \right)
     \tau^a
    \right\} ,
\end{eqnarray}
i.e.
\begin{eqnarray}
    w ( L; \lbrack \chi \rbrack ) &=&
   2 \cos \left(  \frac{ a}{2} \sum_x Q_x ({\bar u}
    + \chi_x ) \right)
 =   2 \cos \left(  \frac{ a}{2} \sum_x Q_x  \chi_x  \right) ,
\end{eqnarray}
with the source
\begin{eqnarray}
   Q_x &=& \delta_{x0} \delta_{y0} ( \delta_{zL} - \delta_{z0} ) .
\end{eqnarray}
The free energy of the static pair is defined by the vacuum
 expectation
value of the Wilson operator:
\begin{eqnarray}
   e^{ - {\cal F} (L) }  & = &
    \frac{ \int {\cal D} \chi w ( L; \lbrack \chi \rbrack )
        e^{- \Gamma \lbrack \chi \rbrack  }
         }{
     \int {\cal D} \chi    e^{- \Gamma \lbrack \chi \rbrack  }
           }
\end{eqnarray}
with the approximate form of the effective action:
\begin{eqnarray}
   \Gamma \lbrack \chi \rbrack &=&
     \frac{1}{2} \sum_{xy} \chi_x ( {\cal D}^{-1} )_{xy} \chi_y .
\end{eqnarray}
The Gaussian integrals can be performed easily:
\begin{eqnarray}
   e^{ - {\cal F} (L) } &=& 2 e^{  - \frac{1}{8} \sum_{xy} Q_x
          {\cal D}_{xy} Q_y              }    .
\end{eqnarray}
 Then we express
the propagator in coordinate space through its Fourier-transform
 and take the sums over $x$ and $y$. This results in
the following expression for the free energy:
\begin{eqnarray}
\label{free}
  {\cal F} (L) &=&  \frac{1}{2} \frac{T}{a} a^3 \int_B
     \frac{ d^3 p}{ (2\pi )^3 } \frac{ \sin^2
          \left(  \frac{1}{2}  p_z L \right)        }{
    {\tilde {\cal D}}^{-1} ( {\vec p}_\perp , p_z ; p^0 =0 )
                                     }
\end{eqnarray}
where the 3-momentum integral is taken over the Brillouin zone.

Calculating the string tension
 we replaced the propagator $\cal D$ by ${\cal D}_0$, i.e.
 neglected the quartic self-interaction
vanishing for $\lambda =1$.

The momentum integral can be performed noticing that the integrand
exhibits poles on the complex $p_z$ plane for vanishing transverse
momentum ${\vec p}_\perp =0$. Therefore it is a good approximation
to
set ${\vec p}_\perp =0$:
\begin{eqnarray}
  {\cal F} (L) &=& \frac{T}{4\pi a} {\cal I} ,
\end{eqnarray}
where the remaining one-dimensional integral,
\begin{eqnarray}
   {\cal I} &=& f_2 a \int_{ - \pi /a}^{\pi /a}
   dp_z \frac{  \sin^2 \left( \frac{1}{2} p_z L \right)    }{
       1 -\lambda +s -s \cos (ap_z )  }
    = \frac{f_2}{s} \int_{-\pi }^\pi dv
   \frac{  \sin^2 \left( \frac{1}{2} \frac{L}{a} v \right)  }{
       \zeta - \cos v }
\end{eqnarray}
with $v = ap_z$, $\zeta = 1 \; + \; (1-\lambda )/s = 1 \; +
 \; m^2 a^2 /2 \ge 1$,
has significantly different properties for $\lambda < 1$
 and $\lambda =1$.

Expressing the denominator in terms of exponentials, the momentum
integral
takes the form ${\cal I} = {\cal I}_0 + {\cal I}_+ + {\cal I}_-$,
where
\begin{eqnarray}
  {\cal I}_0 = \frac{f_2}{2s} \int_{-\pi}^\pi \frac{ dv}{ \zeta -
\cos v }  ,  \qquad
  {\cal I}_\pm = - \frac{f_2}{4s} \int_{-\pi}^\pi \frac{ dv}{
   \zeta -
\cos v } e^{ \pm {\rm i} \frac{L}{a} v } .
\end{eqnarray}
The integrand has simple poles on the complex $v$ plane at $v =
 {\rm
i} v_\pm = {\rm i} \ln ( \zeta \pm \sqrt{ \zeta^2 -1 } )$ $(v_+ >0,
\; v_- < 0)$ corresponding to the zeros of the equation $\zeta
= \cos {\rm i} v_\pm = \cosh v_\pm$.  We calculate these integrals
 by
closing the integration path through the lines
 $C_1$: $(\pi + {\rm i}
0) \to ( \pi + {\rm i} R)$, $C_2$: $(\pi + {\rm i} R) \to (- \pi +
{\rm i} R )$, $C_3$: $( -\pi + {\rm i} R) \to (-\pi + {\rm i} 0)$
for ${\cal I}_0$ and ${\cal I}_+$, and through
 $C_1 '$: $(\pi - {\rm i}
0) \to ( \pi - {\rm i} R)$, $C_2 '$: $(\pi - {\rm i} R) \to (- \pi -
{\rm i} R )$, $C_3 '$: $( -\pi - {\rm i} R) \to (-\pi - {\rm i} 0)$
for  ${\cal I}_-$, where $R \to \infty$.  We show that the
contributions of these lines to the integral vanish for $L/a
=$integer, i.e. in the lattice regularized theory:
\begin{eqnarray}
  \frac{ f_2}{2s} \left( \int_{C_1} + \int_{C_3} \right)
  \frac{ dv}{ \zeta - \cos v }  &=&
     \frac{f_2}{2s} \left(  \int_0^\infty - \int_0^\infty \right)
  \frac{ dv}{ \zeta - \cos v }  = 0 ,
\end{eqnarray}
and using $\cos ( \mp \pi \pm {\rm i} \alpha ) = - \cosh \alpha$,
\begin{eqnarray}
  - \frac{f_2}{4s} \left( \int_{C_1} + \int_{C_2} \right)
   \frac{ dv}{ \zeta - \cos v } e^{ {\rm i} \frac{L}{a} v }
  = - \frac{f_2}{4s} \int_0^\infty \frac{ {\rm i} d\alpha }{
     \zeta + \cosh \alpha }  e^{ - \frac{L}{a} \alpha }
     2 {\rm i} \sin \left( \frac{L}{a} \pi \right)   =0
\end{eqnarray}
for $L/a =$integer. Similarly the contributions from $C_1 '$ and
$C_3 '$ cancel for ${\cal I}_-$, and
\begin{eqnarray}
   \frac{f_2}{2s} \int_{C_2} \frac{ dv}{ \zeta - \cos v} \sim e^{-R}
\to 0 , \nonumber\\
   - \frac{f_2}{4s} \int_{C_2} \frac{ dv}{ \zeta - \cos v} e^{ {\rm
i}  \frac{L}{a} v }  \sim e^{ - \frac{L}{a} R - R } \to 0 ,
         \nonumber\\
   - \frac{f_2}{4s} \int_{C_2 '} \frac{ dv}{ \zeta - \cos v}
    e^{ - {\rm
i}  \frac{L}{a} v }  \sim e^{ - \frac{L}{a} R - R } \to 0
\end{eqnarray}
for $R \to \infty$.
Then we can deform the integration path into the  circles
$C_+$, and $C_-$ of radius $\rho \to 0$ centered at the pole
${\rm i} v_+$ for ${\cal I}_0$ and ${\cal I}_+$, and at
 ${\rm i} v_-$
for ${\cal I}_-$, resp.:
\begin{eqnarray}
\label{cirint}
  {\cal I}_0 &=& \frac{f_2}{2s} \oint_{C_+} \frac{ dv}{ \zeta - \cos
     v } ,
                \nonumber\\
  {\cal I}_+ &=& - \frac{f_2}{4s} \oint_{C_+} \frac{ dv}{ \zeta
   - \cos
     v }     e^{ {\rm i} \frac{L}{a} v  }  ,
                   \nonumber\\
  {\cal I}_- &=&
 - \frac{f_2}{4s} \oint_{C_-} \frac{ dv}{ \zeta - \cos
v }  e^{ -{\rm i} \frac{L}{a} v  } .
\end{eqnarray}
We write $v = {\rm i} v_\pm + \rho e^{ {\rm i} \varphi }$, and
$\zeta - \cos v = {\rm i} \sinh v_\pm  \; \rho e^{ {\rm i} \varphi }
+ \frac{1}{2} \zeta \rho^2 e^{ 2 {\rm i} \varphi }  +
 {\cal O} ( \rho^3 )  $.

Now we have to consider the cases $\lambda < 1$ and $\lambda =1$
independently.
\begin{itemize}
\item For $\lambda < 1$ we have $\zeta >1$ and $v_\pm \neq 0$, and
the nominator is of the order $\rho$. Retaining in the denominator
the terms of the order $\rho$, i.e. writing $\exp \left\{ \pm {\rm i}
\frac{L}{a} \rho e^{ {\rm i} \varphi }  \right\} \approx 1$,
 we get:
\begin{eqnarray}
  {\cal I}_0 &=& \frac{f_2}{2s} \int_0^{2\pi } \frac{ \rho {\rm i}
   d\varphi e^{ {\rm i} \varphi }     }{
  {\rm i} \sinh v_+ \; \rho e^{ {\rm i} \varphi }   }
    = \frac{f_2}{s} \frac{ \pi}{ \sinh v_+ } ,
               \nonumber\\
  {\cal I}_+ &=& - \frac{f_2}{4s} \int_0^{2\pi }
    \frac{ \rho {\rm i} d \varphi \; e^{ {\rm i} \varphi }  }{
            {\rm i} \sinh v_+ \; \rho e^{ {\rm i} \varphi }   }
    e^{ - \frac{L}{a} v_+ }
       = - \frac{f_2}{2s} \frac{ \pi}{ \sinh v_+}
         e^{ - \frac{ L}{a} v_+ } ,
                \nonumber\\
  {\cal I}_- &=&  \frac{f_2}{4s} \int_0^{2\pi }
    \frac{ \rho {\rm i} d \varphi \; e^{ {\rm i} \varphi }  }{
            {\rm i} \sinh v_- \; \rho e^{ {\rm i} \varphi }   }
    e^{  \frac{L}{a} v_- }
       =  \frac{f_2}{2s} \frac{ \pi}{ \sinh v_-}  e^{  \frac{ L}{a}
         v_- } .
\end{eqnarray}
The free energy of the static quark-antiquark pair is then given by:
\begin{eqnarray}
  {\cal F} (L) &=& \frac{T}{4\pi a} \frac{\pi f_2}{s}
   \left( \frac{ 1}{ \sinh v_+}
         - \frac{1}{2 \sinh v_+ } e^{ - \frac{L}{a} v_+ }
         + \frac{1}{2 \sinh v_- } e^{  \frac{L}{a} v_-  } \right) .
\end{eqnarray}
There is an attraction between the quark and the antiquark, which
tends exponentially to zero for increasing separation distance $L$.
A kind of Debye screening takes place governed by the rest mass $m$
of the $A_0$ gluons through $v_\pm$ and $\zeta$.
Consequently the string tension vanishes:
\begin{eqnarray}
    \kappa &=& \lim_{ L \to \infty} \frac{  {\cal F} (L) }{LT}
             = 0.
\end{eqnarray}

\item
For $\lambda =1$ we have $\zeta =1$ and $v_\pm =0$, and the
nominator
 of the integrand on the r.h.s. of the integrals Eq. (\ref{cirint})
becomes of the order $\rho^2$.    The simple poles at $v = {\rm i}
v_\pm$ merged into a single double pole at $v =0$ for
 $\lambda \to 1$.
 Then we have to retain the terms up to
the same order $\rho^2$ in the denominator, writing
$ \exp \left\{ \pm {\rm i} \frac{L}{a} \rho e^{ {\rm i} \varphi }
       \right\} \approx 1 \pm {\rm i} \frac{L}{a} \rho e^{ {\rm i}
  \varphi } $. The terms of the order $1/\rho$ vanish, because they
contain the integral of the function $e^{ - {\rm i} \varphi }$ for a
period. Therefore ${\cal I}_0 =0$ and we get for the free energy:
\begin{eqnarray}
  {\cal F} (L) &=& \frac{T}{4\pi a} \left( {\cal I}_+  + {\cal I}_-
\right)  = \frac{TL}{4a^2} \frac{f_2}{s} .
\end{eqnarray}
We see that the free energy exhibits the `area law'.
Now we obtain a non-vanishing string tension:
\begin{eqnarray}
   \kappa = a^{-2} \frac{f_2}{4s} = a^{-2} \frac{g^2}{8}
          = a^{-2} \frac{-f_4}{8f_2} ,
\end{eqnarray}
where we used $\lambda =1$, i.e. $-f_4 = f_2 g^2$ in the last
equation. The simple proportionality of the string tension $\kappa$
to the strong coupling $g^2$ will be modified
by taking the quartic
self-interaction as perturbation into account.
\end{itemize}

It is interesting to note that the 1--loop effective potential
(\ref{epot1}) becomes constant (the coefficient of the quartic
self-interaction term vanishes)
 for the limit $m \to 0$ which means
that the vacuum state does not break the global center symmetry.
Oppositely, in the case $\lambda < 1$  the effective potential has
two
minima and the vacuum expectation value $\langle u \rangle$ must
 be one of
them, so that the vacuum state violates global center symmetry.
The result obtained is an indication rather than a proof for the
existence of the confining phase in the model, because the 1--loop
approximation can be problematic for $\lambda \approx 1$.
Anyhow, the `area law' of the Wilson loop is a consequence of
the treatment which does not violate center symmetry. Therefore
 we see
that global center symmetry is of extraordinary importance for
the existence
of a phase with non-vanishing string tension even in the continuum
theory.

\section{CONCLUSIONS}
\setcounter{equation}{0}

Summarizing we  constructed the skelet model for $SU(2)$ Yang-Mills
theory with global center symmetry retaining only the $A_0^3$
component of the vector potential.
Due to the treatment not violating global center symmetry
a cut-off dependent  periodic  effective potential occurs
in the skelet model. The minima of this potential are fixed
by charge
conjugation and global center symmetry.
 It has been shown that the $A_0$
gluon vacuum is equivalent with the 4--dimensional Coulomb gas.
This structure of the vacuum
is entirely induced by global center symmetry.

The effective action
for the skelet model has been derived in 1--loop approximation.
In this approximation the effective potential contains only a finite
number of parameters which are particular combinations of the strong
coupling constant and the fugacities of the static charges.
The parameter $\lambda$ is found which governs the loop expansion.
Depending on its value,
the vacuum can be in either deconfined or
confined phase.

For $\lambda < 1$ the string tension vanishes, and the renormalized
 rest
mass of the $A_0$ gluons given by Eq. (\ref{mass1}) is  positive.
Consequently a test $SU(2)$ charge is screened at a distance $1 / m$
due to Debye-screening. The effective potential is periodic and its
curvature at the minima is defined by the renormalized mass $m$.
The ground state is characterized by the non-vanishing
expectation value of the
gluon field (either $a {\bar u} = \pi$ or $-\pi$)
 and the global center symmetry is broken. Thus the
theory decribes the deconfined phase for $\lambda < 1$.

It is
a hint on the possibility of
the existence of a phase with  unbroken global
center symmetry that
the renormalized rest mass  vanishes for $\lambda =1$ in 1-loop
approximation. Then
the           simple poles of the  effective
 propagator of $A_0$ gluons on the complex momentum plane merge
into a single double pole and this leads to a non-vanishing string
tension.
The test charges are not screened now  and the effective
potential becomes a constant. Therefore the global
center symmetry broken for $\lambda <1$ is restored for $\lambda =1$
in the ground state. These are just the properties of the
confined phase.

Finally some remarks on our procedure and the results obtained above:

\begin{itemize}
\item
Our procedure of deriving the effective theory contains a
non-perturbative resummation of  infinitely many vertices generated
by the periodic effective potential, i.e. by global center symmetry.
This results in a vacuum structure being equivalent with a
4--dimensional Coulomb gas.
Due to
reserving global center symmetry during the whole treatment we
expect our approach to reveal the IR (long distance) features
 of $SU(2)$
Yang-Mills theory correctly. On the other hand, it is necessary to
incorporate the
spatial  components of the gauge field for the extension of
 the model to
the UV regime.
\item
Our gauge fixing does not violate global center symmetry and is done
rather consistently than in
 our earlier treatment\cite{Sai93}.  We have shown that
 the periodic effective potential is completely determined by the
Haar measure. It is independent of the explicit form of
the periodic gauge condition $F^3$, as
 one   expects\cite{Pol88,Pol91}.

\item Part of the self-interaction (higher order in the field) and
the terms higher order in the loop expansion were neglected.
The qualitative features of the effective potential determined by
global center symmetry are not sensitive to this approximation. The
critical value of the parameter $\lambda$ for which confinement sets
on can be modified by the self-interaction and higher order terms in
the loop expansion.
\end{itemize}

\section*{ACKNOWLEDGEMENT}

The author is deeply indebted to J. Pol\'onyi and A. Sch\"afer for
the stimulating and useful discussions, and to W.
Greiner for his permanent interest and kind hospitality in
 Frankfurt.
This work has been supported by a joint project of the
 Hungarian Academy
of Sciences and the Deutsche Forschungsgemeinschaft,
by the EC project ERB-CIPA-CT93-0651 (Prop. 3560) and by the
Hungarian Research Fund (OTKA 2192/91).

\appendix
\def\thesection{\Alph{section}}
\def\theequation{\Alph{section}.\arabic{equation}}

\section{Vacuum to vacuum transition amplitude}
\setcounter{equation}{0}

Let us denote by ${\hat H}$ the Hamiltonian of $SU(2)$ Yang-Mills
 theory
in the gauge $A_0^a \equiv 0$ $(a=1,2,3)$.
Dividing the interval $T$ of the whole time evolution into $N$
intervals of length $a$ and inserting the projector operator
$P_\theta$ of the colour neutral physical states at each time slice,
we write the physical vacuum to vacuum transition amplitude
as the product of infinitesimal physical transition amplitudes:
\begin{eqnarray}
\lefteqn{
  \langle 0 | e^{ - {\rm i} {\hat H} T} | 0 \rangle_ {\mbox {phys}}
   =    }
        \nonumber\\
    & &
   \left( \prod_{i =1}^{N-1}  \sum_{ k_i =0,1 }
   \prod_{\vec x}   \int dA^{ja}_{ {\vec x} t_i}
    \int_{S_{2\pi}} d_H \left( \frac{ ga \alpha^a_{ {\vec x} t_i}  }{
             T}  \right)   \right)
   \prod_{k=0}^{N-1}
    \left\langle A^{h( \alpha ) \; ja}_{ {\vec x} t_{k +1} }
   \right| e^{ - {\rm i} a {\hat H}  }  \left|
    A^{ja}_{ {\vec x} t_k }  \right\rangle ,
\end{eqnarray}
where $h (\alpha_{ {\vec x}t_i  } ) = h_0 (\alpha_{ {\vec x} t_i } )
 z_1^{k_i} $, $h_0 \in SU(2)$.
Using the explicit form of the infinitesimal transition
amplitude\cite{Pol92},
\begin{eqnarray}
    \left\langle A^{h( \alpha ) \; ja}_{ {\vec x} t_{k +1} }
   \right| e^{ - {\rm i} a {\hat H}  }  \left|
    A^{ja}_{ {\vec x} t_k }  \right\rangle
   &=& \left. \exp \left\{  - {\rm i} a \frac{1}{4g^2}
     \int d^3 y {\bar F}_{\mu \nu}^a {\bar F}^{\mu \nu \; a}
             \right\}  \right|_{ t=t_k} ,
\end{eqnarray}
and reintroducing the zeroth component of the vector potential by
Eq. (\ref{A0}), we obtain:
\begin{eqnarray}
  \langle 0 | e^{ - {\rm i} {\hat H} T} | 0 \rangle_ {\mbox {phys}}
      &=&
 \int_{\cal G} {\cal D }_H (a A^{0a} ) \int {\cal D} A^{ja}
    e^{  {\rm i} S \lbrack A^{ h(A_0 ) \; j} , A^j \rbrack   }  .
\end{eqnarray}

Now we introduce the gauge conditions $F^a$ by inserting
1 in the form of (\ref{unity}). Dividing by the multiple of the
group volume, we get:
\begin{eqnarray}
  \langle 0 | e^{ - {\rm i} {\hat H} T} | 0 \rangle_ {\mbox {phys}}
 =
\frac{ \int_{ \cal G} {\cal D}_H ( a A^{0a} )
   \int {\cal D} A^{ia} \Delta \lbrack A \rbrack
   \int_{\cal G} {\cal D} h' \left( \prod_{a=1}^3
   \delta \lbrack F^a \lbrack A^{h'} \rbrack \rbrack \right)
   e^{ {\rm i} S \lbrack A^{ h (A_0 ) \; j} , A^j \rbrack   }
     }{
  \int_{\cal G} {\cal D} h'
      }  .
     \nonumber\\
\end{eqnarray}
Making use of the fact that the integration measure, the functional
$\Delta \lbrack A \rbrack$, and the action are invariant under any
gauge transformation $h'$, we can write the denominator in the
following form:
\begin{eqnarray}
  \int_{\cal G} {\cal D} h'
 \int_{ \cal G} {\cal D}_H \left( a A^{h' \;0a} \right)
   \int {\cal D} A^{h' \; ia} \Delta \lbrack A^{h'} \rbrack
   \left( \prod_{a=1}^3
   \delta \lbrack F^a \lbrack A^{h'} \rbrack \rbrack \right)
   e^{ {\rm i} S \lbrack ( A^{ h (A_0 )} )^{h' \; j} ,
    A^{h' \; j} \rbrack   } .
\end{eqnarray}
Further on we make use of the transformation law of the parameters
 of
the gauge transformations (see Eq. (\ref{trafo}) ):
\begin{eqnarray}
    h(A_0 ) h' &=& h ( A_0^{h'} ) .
\end{eqnarray}
Then we can remove $h'$ from the integration variables $A^{h'\; 0}$,
and $A^{h'\; j}$:
\begin{eqnarray}
\label{vacvac}
  \langle 0 | e^{ - {\rm i} {\hat H} T} | 0 \rangle_ {\mbox {phys}}
  &=&
 \int_{ \cal G} {\cal D}_H \left( a A^{0a} \right)
   \int {\cal D} A^{ ia} \Delta \lbrack A \rbrack
   \left( \prod_{a=1}^3
   \delta \lbrack F^a \lbrack A \rbrack \rbrack \right)
   e^{ {\rm i} S \lbrack  A^{ h (A_0 ) \; j} , A^j \rbrack   }
  .
     \nonumber\\
\end{eqnarray}

\section{$\Delta \lbrack A \rbrack$ functional}
\setcounter{equation}{0}

Let us write the general gauge transformation $h_{{\vec x} t} \in
SU(2)$  in the form:
\begin{eqnarray}
  h_{ {\vec x} t} = h_0 (\beta^a_{{\vec x}t} ) z_1^{n_t} ,
\end{eqnarray}
where $a \beta_{ {\vec x}t} \in \lbrack 0, 2\pi \rbrack$, and
$ n_t =0,1$ integer at the time slice $t$. (For the sake of
 simplicity we
omit the lower indeces $({\vec x}, t)$ of the variables if it is
not confusing.)

According to Eq. (\ref{trafo}),
the following general expressions are obtained:
\begin{eqnarray}
\label{A0h0}
   ( A_0^{h_0} )^1 \frac{  \sin \left( \frac{1}{2} a A_0^{h_0}
                       \right)  }{    A_0^{h_0}         }
   &=& \left( \beta^1 \cos \frac{ a A_0}{ 2}  + \beta^2
                      \sin \frac{ a A_0}{ 2}  \right)
       \frac{ \sin \left( \frac{1}{2} a \beta \right)  }{ \beta }  ,
                   \nonumber\\
   ( A_0^{h_0} )^2 \frac{  \sin \left( \frac{1}{2}  a A_0^{h_0}
                                   \right) }{   A_0^{h_0}     }
   &=& \left( \beta^2 \cos \frac{ a A_0}{ 2}  - \beta^1
                      \sin \frac{ a A_0}{ 2}  \right)
       \frac{ \sin \left( \frac{1}{2} a \beta \right) }{ \beta }  ,
                   \nonumber\\
   ( A_0^{h_0} )^3 \frac{  \sin \left( \frac{1}{2} a A_0^{h_0}
              \right)  }{   A_0^{h_0}              }
   &=& \frac{\beta^3}{\beta} \cos \frac{ a A_0}{ 2}
        \sin \frac{ a \beta}{2}
     + \frac{ A_0^3}{A_0} \sin \frac{ a A_0}{2} \cos \frac{ a
                \beta}{2}  ,
\end{eqnarray}
and
\begin{eqnarray}
   ( A_0^{h} )^1 &=& \frac{ A_0^{h_0} (-1)^{n_t} + 2\pi n_t /a  }{
          \sin \left( \frac{1}{2} \alpha_{n_t} \right)   }
    \left\lbrack
    \beta^1 \cos \left( \frac{1}{2} \alpha_{n_t} \right)
  + \beta^2 \sin \left( \frac{1}{2} \alpha_{n_t} \right)
    \right\rbrack
    \frac{ \sin \left( \frac{1}{2} a \beta \right)  }{ \beta }  ,
                   \nonumber\\
   ( A_0^{h} )^2 &=& \frac{ A_0^{h_0} (-1)^{n_t} + 2\pi n_t /a  }{
           \sin \left( \frac{1}{2} \alpha_{n_t} \right)  }
    \left\lbrack
    \beta^2 \cos \left( \frac{1}{2}  \alpha_{n_t} \right)
  - \beta^1 \sin \left( \frac{1}{2}  \alpha_{n_t} \right)
    \right\rbrack
    \frac{ \sin \left( \frac{1}{2} a \beta \right)  }{ \beta }  ,
                   \nonumber\\
   ( A_0^{h} )^3 &=&  \frac{  A_0^{h_0} (-1)^{n_t} + 2\pi n_t  }{
          \sin \left( \frac{1}{2} \alpha^{n_t} \right)  }
   \left\lbrack
   \frac{\beta^3}{\beta} \cos \left( \frac{1}{2} \alpha_{n_t}
                              \right)  \sin \frac{ a \beta}{2}
                   \right.           \nonumber\\
          &  &  \left.
 + (-1)^{n_t} \frac{ A_0^3 }{ A_0 }
       \sin \left( \frac{1}{2} \alpha_{n_t} \right)
       \cos \frac{ a \beta}{2}
   \right\rbrack  ,
\end{eqnarray}
with $\alpha_{n_t} = aA_0 (-1)^{n_t} $, where we made use of the
 equations:
\begin{eqnarray}
 (aA_0^a )^{z_1^{n_t} } &=& aA_0^a + 2\pi n_t (-1)^{n_t}
    \frac{aA_0^a}{ aA_0}  ,  \nonumber\\
 \sin \left( \frac{1}{2} aA_0^{z_1^{n_t} } \right)
  & =& (-1)^{n_t} \sin \left( \frac{1}{2} \alpha_{n_t} \right) ,
                \nonumber\\
 \cos \left( \frac{1}{2} aA_0^{z_1^{n_t} } \right)
  & = & (-1)^{n_t} \cos \left( \frac{1}{2} \alpha_{n_t} \right) .
\end{eqnarray}
The expressions for
 $\sin \left( a A_0^{h_0} /2 \right)$, i.e. for $A_0^{h_0}$
in terms of $\beta^a$'s we find
by taking the sum of the squares of the expressions given by
(\ref{A0h0}):
\begin{eqnarray}
 \sin^2 \frac{ a A_0^{h_0} }{2}
    &=&
  \left\lbrack 1 - (
    1 + \cos^2 \theta ) \sin^2 \frac{a A_0 }{2}
   \right\rbrack
     \sin^2 \left( \frac{1}{2} a \beta \right)
     \nonumber\\
          &  & +   \sin^2 \frac{a A_0}{2}
     + \frac{1}{2}  \cos \theta \frac{A_0^3}{A_0}
                   \sin ( a \beta )    \sin ( a A_0 )
                      \nonumber\\
     & \equiv &   B ( a \beta , \theta ) ,
\end{eqnarray}
where the spherical coordinates
\begin{eqnarray}
   \beta^1 = \beta \sin \theta \cos \varphi ,
   \qquad
   \beta^2 = \beta \sin \theta \sin \varphi ,
    \qquad
    \beta^3 = \beta \cos \theta
\end{eqnarray}
have been introduced. Then we get:
\begin{eqnarray}
  a A_0^{h_0} &=& {\mbox {arccos}} \lbrack 1 - 2B ( a \beta ,
                     \theta )  \rbrack .
\end{eqnarray}

The inverse of the functional $\Delta \lbrack A \rbrack$ is
 given by:
\begin{eqnarray}
  \Delta^{-1} \lbrack A \rbrack &=&
  \sum_{ \{ n_t  \} }  \int {\cal D}_H ( a \beta^a )
      \prod_{a=1}^3 \delta \lbrack F^a \lbrack A^{ h_0 z^{n_t} }
       \rbrack
                           \rbrack
                   \nonumber\\
    &=& \sum_{ \{ n_t \} }  \left( \prod_{ {\vec x} t}
       \int_0^{2\pi}  d( a \beta_{ {\vec x}t }  )  \;
       ( a \beta_{ {\vec x}t} )^2
    \int_0^\pi \sin \theta_{{\vec x}t} \; d\theta_{{\vec x}t}
       \int_0^{2\pi}  d\varphi_{{\vec x}t}
       \frac{1}{4\pi^2}
       \frac{  \sin^2 \left(  \frac{1}{2} a \beta_{{\vec x}t}
        \right) }{
       (a \beta_{{\vec x}t} )^2 }
                             \right)        \cdot
                     \nonumber\\
       & &   \cdot
       \prod_{a=1}^3 \delta  \lbrack F^a \lbrack A^{h_0 z^{n_t} }
                                \rbrack    \rbrack     .
\end{eqnarray}
Here we can write:
\begin{eqnarray}
   \delta \lbrack F^1 \lbrack A^{h_0 z_1^{n_t} }  \rbrack \rbrack
    &=&  \prod_{ {\vec x}t }
   \frac{  \delta ( \varphi - \varphi_\nu )   }{
     \left|
    \frac{ A_0^{h_0} (-1)^{n_t} + 2 \pi n_t /a }{
    \sin \left( \frac{1}{2} a A_0^{h_0} (-1)^{n_t}  \right)     }
    \sin \left( \frac{1}{2} a \beta \right)  \sin \theta
    \sin \left( \frac{1}{2} a A_0 (-1)^{n_t} - \varphi_\nu \right)
     \right|                                                   }
                \nonumber\\
    &=&  \prod_{ {\vec x}t }
    \frac{  \delta ( \varphi - \varphi_\nu )   }{
     \left|
    \frac{ A_0^{h_0} (-1)^{n_t} + 2 \pi n_t /a }{
    \sin \left( \frac{1}{2} a A_0^{h_0} (-1)^{n_t}  \right)     }
    \sin \left( \frac{1}{2} a \beta \right)  \sin \theta
     \right|                                                   }   ,
\end{eqnarray}
where the equation
\begin{eqnarray}
  \cos \left(    \frac{1}{2} a A_0 (-1)^{n_t}  - \varphi_\nu \right)
 &=& 0
\end{eqnarray}
has 1 root for $\varphi_\nu \in \lbrack 0 , 2\pi \rbrack$ and $n_t
=0$ of the
form $ \varphi_\nu = \lbrack a A_0 (-1)^{n_t} - (2\nu +1 )
 \pi \rbrack /2 $
with  $\nu =0$ for $\pi /2 \le a A_0 /2 \le \pi$, and
  $\nu =-1$ for  $0 \le a A_0 /2 \le \pi /2$. For $n_t =1$ we can
change the integration variable from $\varphi$ to $-\varphi$ and
then
we get the same root as for $n_t =0$
(notice that $| \sin ( a A_0 /2  - \varphi_\nu ) | = 1$);
\begin{eqnarray}
   \delta \lbrack F^2 \lbrack A^{h_0 z_1^{n_t} }  \rbrack \rbrack
    &=&  \prod_{ {\vec x}t }
   \sum_{\mu =1,2} \frac{  \delta ( \theta - \theta_\mu )   }{
     \left|  - \frac{\partial }{ \partial \theta }
    \frac{ A_0^{h_0} (-1)^{n_t} + 2 \pi n_t /a }{
    \sin \left( \frac{1}{2} a A_0^{h_0} (-1)^{n_t}  \right)    }
    \sin \left( \frac{1}{2} a \beta \right)  \sin \theta
    \sin \left( \frac{ a A_0}{2} - \varphi_\nu \right)
     \right|                                                   }
                \nonumber\\
    &=&  \prod_{ {\vec x}t }
   \sum_{\mu =1,2} \frac{  \delta ( \theta - \theta_\mu )   }{
     \left|
    \frac{ A_0^{h_0} (-1)^{n_t} + 2 \pi n_t /a }{
    \sin \left( \frac{1}{2} a A_0^{h_0} (-1)^{n_t}  \right)    }
    \sin \left( \frac{1}{2} a \beta \right)  \cos \theta_\mu
     \right|                                                   } ,
                \nonumber\\
    &=&  \prod_{ {\vec x}t }
   \sum_{\mu =1,2} \frac{  \delta ( \theta - \theta_\mu )   }{
     \left|
    \frac{ A_0^{h_0} (-1)^{n_t} + 2 \pi n_t /a }{
    \sin \left( \frac{1}{2} a A_0^{h_0} (-1)^{n_t}  \right)      }
    \sin \left( \frac{1}{2} a \beta \right)
     \right|                                                   } ,
\end{eqnarray}
where $\sin \theta_\mu =0$ has 2 roots, $\theta_\mu = 0, \pi$.
Now we perform the integrals over the angles:
\begin{eqnarray}
\lefteqn{
  \Delta^{-1} \lbrack A \rbrack   =
        }     \nonumber\\
 & =& \sum_{ \{ n_t \} }  \left( \prod_{ {\vec x} t}
       \int_0^{2\pi}  d( a \beta_{ {\vec x}t }  )  \;
       \frac{1}{4 \pi^2}  \frac{1}{2}
          \sum_{\mu }  \left.
 \frac{  \sin^2 \left( \frac{1}{2}   a A_0^{h_0} (-1)^{n_t} \right)
      }{  \left( a A_0^{h_0} (-1)^{n_t} + 2\pi n_t   \right)^2
                                }    \right|_{\theta_\mu }
         \right)          \left.
       \delta  \lbrack F^3 \lbrack ( A^{h_0 z_1^{n_t} } )_0^3
                                \rbrack    \rbrack
                           \right|_{\theta_\mu \varphi_\nu} .
                      \nonumber\\
\end{eqnarray}

For $\theta_\mu =0, \pi$ and $A_0^1 =A_0^2 =0$:
\begin{eqnarray}
  B (a \beta , \theta_\mu =0,\pi )
    &=& \frac{1}{2} \left( 1 - \cos \alpha_\mu
                          \right)
    = \sin^2 \left( \frac{\alpha_\mu }{2}
            \right)
\end{eqnarray}
with $\alpha_\mu = a(A_0^3 + \beta \cos \theta_\mu )$,
and then we obtain:
\begin{eqnarray}
  a A_0^{h_0} &=&
   \left\{  \begin{array}{lll}
     \left| \alpha_\mu \right|
     &  {\mbox {for}} &  | \alpha_\mu  | \in
      \lbrack 0, 2\pi \rbrack                   \\
  2\pi -   \left| \alpha_\mu  \right|
     &  {\mbox {for}} &  | \alpha_\mu  | \in
      \lbrack  2\pi , 4\pi \rbrack
              \end{array}   \right.   ,
\end{eqnarray}
and
\begin{eqnarray}
\label{aa03}
 aA_0^{h_0 \; 3} &=& A_0^{h_0} \frac{
   \sin \left( \frac{\alpha_\mu}{2}  \right)
  }{  \sin \left( \frac{a}{2} A_0^{h_0} \right)    }
      = \left\{   \begin{array}{lll}
    \alpha_\mu  &
     {\mbox {for}}  & | \alpha_\mu | \in
         \lbrack 0, 2\pi \rbrack      \\
   \alpha_\mu  - 2\pi {\mbox {sgn}}
    \alpha_\mu    &
     {\mbox {for}}  & | \alpha_\mu | \in
         \lbrack  2\pi , 4\pi \rbrack
                   \end{array}      \right.
\end{eqnarray}
As the function ${\cal F} (\alpha )$ is invariant under global
 center
transformations,  the sums over $n_t$'s can be carried out, and
the inverse of the $\Delta$ functional takes the following form:
\begin{eqnarray}
  \left.  \Delta^{-1} \lbrack A \rbrack
     \right|_{A_0^1 = A_0^2 =0}
    &=& \sum_{ \{ n_t \} }  \left( \prod_{ {\vec x} t}
       \int_0^{2\pi}  d( a \beta_{ {\vec x}t }  )  \;
       \frac{1}{4 \pi^2}  \frac{1}{2}
          \sum_{\mu }
 \frac{  \sin^2 \left\lbrack \frac{   a }{2}
             \left( A_0^3 + \beta \cos \theta_\mu \right)
                \right\rbrack
                               }{  \left(  (-1)^{n_t}
            \left. a A_0^{h_0}  \right|_{\theta_\mu }
                      + 2\pi n_t   \right)^2
                                }
                   \right. \cdot   \nonumber\\
   & &    \cdot \left. \left.
       \delta  \left(
         \left. {\cal F} ( a A_0^{h_0 \; 3} ) \right|_{t+a }   -
         \left. {\cal F} ( a A_0^{h_0 \; 3} ) \right|_{t }
               \right)
             \right) \right|_{\theta_\mu , \; A_0^1 =A_0^2 =0}
                \nonumber\\
    &=&  \prod_{ {\vec x} t}
       \int_0^{2\pi}  d( a \beta_{ {\vec x}t }  )  \;
       \frac{1}{4 \pi^2}  \frac{1}{2}
          \sum_{\mu }
   \sin^2 \left( \frac{   a }{2}
             \left( A_0^3 + \beta \cos \theta_\mu \right)
                \right)
              \cdot      \nonumber\\
  & & \cdot     \left\lbrack
       \frac{1}{ ( \left. aA_0^{h_0}  \right|_{\theta_\mu
                             } )^2      }
    +  \frac{1}{ ( 2\pi -
                 \left. aA_0^{h_0}  \right|_{\theta_\mu   } )^2  }
       \right\rbrack
                    \cdot   \nonumber\\
   & &    \cdot
      \left.  \delta  \left(
         \left. {\cal F} ( a A_0^{h_0 \; 3} ) \right|_{t+a }   -
         \left. {\cal F} ( a A_0^{h_0 \; 3} ) \right|_{t }
               \right)
             \right|_{\theta_\mu , \; A_0^1 =A_0^2 =0} .
\end{eqnarray}
Inserting the  expression (\ref{aa03}) we find:
\begin{eqnarray}
  \left.  \Delta^{-1} \lbrack A \rbrack  \right|_{A_0^1 = A_0^2 =0}
    &=&  \prod_{ {\vec x} t}
       \int_0^{2\pi}  d( a \beta_{ {\vec x}t }  )  \;
       \frac{1}{4 \pi^2}  \frac{1}{2}
          \sum_{\mu }
   \sin^2 \left( \frac{   a }{2}
             \left( A_0^3 + \beta \cos \theta_\mu \right)
                \right)
             \cdot    \nonumber\\
           & &   \cdot
       \left\lbrack
       \frac{1}{ | a(A_0^3 + \beta \cos \theta_\mu ) |^2  }
    +  \frac{1}{ \left\lbrack 2\pi -
                 | a (A_0^3 + \beta \cos \theta_\mu  ) |
                  \right\rbrack^2
}
       \right\rbrack
                    \cdot   \nonumber\\
   & &    \cdot
       \delta  \left(
         \left. {\cal F} ( a (A_0^3  +
          \beta \cos \theta_\mu )) \right|_{t+a }      -
         \left. {\cal F} ( a (A_0^3  +
          \beta \cos \theta_\mu )) \right|_{t }
               \right)           .
                    \nonumber\\
\end{eqnarray}

We introduce the new integration variable:    $ \xi = a \beta \cos
\theta_\mu $. Then the integrand becomes independent of $\mu$ and
 the sum over $\mu$ can be carried out:
\begin{eqnarray}
   \sum_{\mu} \frac{1}{ \cos \theta_\mu } \int_0^{ 2\pi \cos
\theta_\mu } d \xi  \ldots
   &=& \int_{-2\pi }^{ 2\pi } d \xi \ldots .
\end{eqnarray}
Furthermore
we have to use the field configurations satisfying
${\cal F} (a A_0^3 )_{t + a } - {\cal F} ( a A_0^3 )_t =0$, i.e.
$ \left. a A_0^3 \right|_t = \left. aA_0^3 \right|_{t+a} $, and
$ \left. aA_0^3 \right|_{t+a} - 2\pi {\mbox {sgn}} \left. A_0^3
 \right|_{t+a} $. As ${\cal F}$ is periodic, the Dirac delta
can be rewritten in both cases in  the same  way:
\begin{eqnarray}
  \delta \left( {\cal F}  ( a A_0^3 + \xi )_{t+a }
          - {\cal F} ( a A_0^3 + \xi )_t \right)
             &=&
   \sum_k \frac{  \delta ( \xi_t - \xi_{kt} )   }{
    | {\cal F}' (  a A_{0 \; t}^3 + \xi_t ) |    } ,
\end{eqnarray}
According to the requirements satisfied by  ${\cal F} (\alpha )$,
 the
argument of the Dirac delta has 2 zeros in the interval $\xi_{kt}
\in \lbrack -2\pi , 2 \pi \rbrack$: $\xi_{kt} =
\xi_{t+a}$, $\xi_{t+a} - 2\pi {\mbox {sgn}} \xi_{t+a}$.
For both zeros hold the equations
\begin{eqnarray}
  \sin^2 \left( \frac{1}{2} ( a A_{0 \; t}^3 + \xi_{kt} ) \right)
 &=&  \sin^2 \left( \frac{1}{2} ( a A_{0 \; t+a}^3 + \xi_{t+a} )
  \right) ,
       \nonumber\\
 \left|  {\cal F}' ( a A_{0 \; t}^3 + \xi_{ kt} ) \right| &=&
 \left|  {\cal F}' ( a A_{0 \; t+a}^3 + \xi_{t+a} ) \right| .
\end{eqnarray}
In the case of the second zero we can make the integral
transformation from $\xi_{t+a}$ to $ \xi_{t+a} - 2\pi {\mbox {sgn}}
\xi_{t+a}$, under which the rest of the integrand remains unaltered.
Then we can take for both zeros:
\begin{eqnarray}
\lefteqn{
 \frac{1}{ | aA_{0 \; t+a}^3 + \xi_{kt} |^2 } +
 \frac{1}{ \left( 2\pi - | aA_{0 \; t+a}^3 + \xi_{kt} | \right)^2 }
                \to
           }    \nonumber\\
  & \to &
 \frac{1}{ | aA_{0 \; t+a}^3 + \xi_{t+a} |^2 } +
 \frac{1}{ \left( 2\pi - | aA_{0 \; t+a}^3 + \xi_{t+a} | \right)^2 }
{}.
\end{eqnarray}
Finally both zeros give the same contribution to the integral
cancelling the factor $\frac{1}{2}$ before the sum over $\mu$.
We integrate over the $\beta_{ {\vec x} t}$'s successively according
to their time argument $t$ simply writing $\xi_{kt} = \xi_{t + a}$:
\begin{eqnarray}
 \left. \Delta^{-1} \lbrack A \rbrack     \right|_{F \equiv 0}
              &=&
 \prod_{ {\vec x}t } \left\lbrack \frac{ 1}{ 4\pi^2}
            \frac{   \sin^2 \left( \frac{1}{2}
                ( a A_{0 \; {\vec x}t}^3  + \xi_{ {\vec x} T/2} )
                                \right)   }{
                 | {\cal F}' ( a A_{0 \; {\vec x}t}
                  + \xi_{ {\vec x}T/2
                      } )    |               }     \cdot
           \right.     \nonumber\\
      & & \left.  \cdot  \left(
    \frac{ 1}{ \left| a A_{0 \; {\vec x}t}^3 +
                \xi_{ {\vec x}T/2} \right|^2       }
   + \frac{ 1}{ \left(2\pi - \left| a A_{0 \; {\vec x}t}^3 +
                \xi_{ {\vec x}T/2} \right|  \right)^2       }
                  \right)
          \right\rbrack .
\end{eqnarray}
Choosing   $\xi_{ {\vec x} T/2} \to 0$
 for the pure gauge final state,
 we obtain:
\begin{eqnarray}
 \left. \Delta^{-1} \lbrack A \rbrack     \right|_{F \equiv 0}
              &=&
 \prod_{ {\vec x}t } \left\lbrack \frac{ 1}{4 \pi^2}
            \frac{   \sin^2 \left( \frac{1}{2}
                 a A_{0 \; {\vec x}t}^3
                                \right)   }{
                 | {\cal F}' ( a A_{0 \; {\vec x}t}
                        )    |               }
      \left(
   \frac{ 1}{ \left( a A_{0 \; {\vec x}t}^3
                  \right)^2       }    +
   \frac{ 1}{ \left( 2\pi - | a A_{0 \; {\vec x}t}^3 |
                  \right)^2       }
      \right)
                    \right\rbrack .
                \nonumber\\
\end{eqnarray}

\section{Explicit form of the vacuum to vacuum transition amplitude}
\setcounter{equation}{0}

We have made  the jumps of the vector potential $A_0$ explicit
 in the
path integral. This resulted in the sum over center equivalent
copies
of any field configuration. As the Haar measure, the $\Delta$
functional,  the action, and the gauge condition $F^3$ are invariant
 under
global center transformations, only the arguments of the Dirac
 deltas
with the gauge conditions $F^1$ and $F^2$ transform in a non-trivial
way under global center transformation. Taking this into account, we
obtain:
\begin{eqnarray}
\lefteqn{
   {\cal Z} \lbrack 0 \rbrack =
       }      \nonumber\\
   &=&
   \sum_{  \{ n_t \} } \int_{ S_{2\pi }  }
    {\cal D}_H ( a A_0^a ) \int {\cal D} A_{ia}
   \Delta \lbrack A \rbrack_{ F \equiv 0 }
   \left( \prod_{a=1}^3 \delta \left\lbrack F^a \left\lbrack
     A_0^a \left( 1 + \frac{ 2\pi n_t (-1)^{n_t} }{ a A_0 } \right) ,
     A_i^a \right\rbrack \right\rbrack     \right)  \cdot
                 \nonumber\\
   & & \cdot \left. e^{ {\rm i} S \lbrack A^{ h_0 (A_0 )  \; j} ,
                                   A^j  \rbrack              }
                    \right|_{ F \equiv 0}
                   \nonumber\\
   &=& \sum_{ \{ n_t \} }  \int {\cal D}_{ {\mbox {flat}} }
      ( a A_0^a )  \int {\cal D} A_i^a
     \frac{  \exp \left\{ \sum_x \ln \sin^2 \left( \frac{1}{2} a A_0
                                            \right)
                            - \sum_x \ln ( 4\pi^2 )  \right\}
             }{  \prod_x ( a A_0 + 2\pi n_t (-1)^{n_t} )^2  }
    \Delta \lbrack A \rbrack_{ F \equiv 0 }  \cdot
                \nonumber\\
    & & \cdot \left\lbrack \prod_x
      \delta \left( a A_0^1 \left( 1 + \frac{2\pi n_t
       (-1)^{n_t} }{ a A_0
                                      } \right)   \right)
      \delta \left( a A_0^2 \left( 1 + \frac{2\pi n_t
       (-1)^{n_t} }{ a A_0
                                      } \right)   \right)
                    \right.   \cdot
                \nonumber\\
    & & \cdot \left.
      \delta \left( {\cal F} ( a A_0^3 )_{t+a }
                   - {\cal F} ( a A_0^3 )_t   \right)
              \right\rbrack
      \left. e^{  {\rm i} S \lbrack A^{ h_0 (A_0 ) \; j} , A^j
                                     \rbrack                    }
               \right|_{ F \equiv 0 }
                         \nonumber\\
   &=& \sum_{ \{ n_t \} }  \int {\cal D}_{ {\mbox {flat}} }
      ( a A_0^a )  \int {\cal D} A_i^a
     \frac{  \exp \left\{ \sum_x \ln \sin^2 \left( \frac{1}{2} a A_0
                                            \right)
                            - \sum_x \ln ( 4\pi^2 )  \right\}
             }{  \prod_x ( a A_0
             + 2 \pi n_t (-1)^{n_t} )^2  }
      \cdot       \nonumber\\
    & & \cdot \exp \left\{ \sum_x \ln (4 \pi^2 )
              - \sum_x \ln \sin^2 \left( \frac{1}{2} a A_0 \right)
             + \sum_x \ln | {\cal F}' (a A_0^3 ) |
                   \right\}     \cdot
                \nonumber\\
     & & \cdot  \left\lbrack \frac{1}{ (aA_0 )^2 }
       + \frac{1}{ ( 2\pi - aA_0 )^2  }   \right\rbrack^{-1}
         \cdot         \nonumber\\
    & & \cdot \left( \prod_x
      \delta \left( a A_0^1    \right)
      \delta \left( a A_0^2    \right)
      \delta \left( {\cal F} ( a A_0^3 )_{t+a }
                   - {\cal F} ( a A_0^3 )_t   \right)
              \right)
     \left. e^{  {\rm i} S \lbrack A^{ h_0 (A_0 ) \; j} , A^j
                                     \rbrack                    }
               \right|_{ F \equiv 0 }  .
\end{eqnarray}
We perform now the sums over $n_t$'s:
\begin{eqnarray}
   {\cal Z} \lbrack 0 \rbrack
   &=&
   \int {\cal D}_{\mbox {flat}} A_0^3 \int {\cal D} A_i^a
   \left( \prod_x \delta \left( {\cal F} ( aA_0^3
    )_{{\vec x} \; t+a}
       - {\cal F} ( aA_0^3 )_{ {\vec x}t} \right) \right)
            \cdot        \nonumber\\
    & &  \cdot
     \left. e^{  {\rm i} S \lbrack A^{ h_0 (A_0 ) \; j} , A^j
       \rbrack
             + \sum_x \ln | {\cal F}' (aA_0^3 )  |
                            }
               \right|_{ F \equiv 0 }  .
\end{eqnarray}
 In order to recover
the effective potential induced by the Haar measure,
 we rewrite the Dirac delta on the r.h.s.
  and show that the
physical vacuum to vacuum amplitude is independent of the choice
of the gauge function ${\cal F} (\alpha )$. The argument of the
 Dirac
delta exhibits 2 zeros: $aA_{0 \; t}^3 = a A_{0 \; t+a}^3$, and
$ a A_{0 \; t+a}^3  - 2\pi {\mbox {sgn}} A_{0 \; t}^3     $.
Consequently, we can write:
\begin{eqnarray}
\lefteqn{
  \delta \left( {\cal F} ( aA_0^3 )_{t+a}  -
                {\cal F} ( aA_0^3 )_t        \right)  =
        }     \nonumber\\
   &=&
   \frac{1}{ | {\cal F}' (aA_{0 \; t}^3  )  |  }
   \left\lbrack  \delta \left( aA_{0 \; t}^3 - aA_{0 \; t+a}^3
                        \right)
               +
                 \delta \left( aA_{0 \; t}^3 - aA_{0 \; t+a}^3
                        + 2 \pi {\mbox {sgn}} A_{0 \; t+a}^3
                         \right)
   \right\rbrack .
\end{eqnarray}
We make the
integral transformation  $A_{0 \; t+a}^3 \to
A_{0 \; t+a}^3 - 2\pi {\mbox {sgn}} A_{0 \; t+a}^3 $ in the second
 term.
 Under this
transformation the rest of the integrand remains unaltered as
${\cal F}$, ${\cal F}' $, and the action $S$ remain invariant.
Then both terms give the same contribution to the path integral
and we can replace the Dirac delta as follows:
\begin{eqnarray}
  \delta \left( {\cal F} ( aA_0^3 )_{t+a}  -
                {\cal F} ( aA_0^3 )_t        \right)
   &  {\rightarrow } &
   \frac{2}{ | {\cal F}' (aA_{0 \; t}^3  )  |  }
     \delta \left( aA_{0 \; t}^3 - aA_{0 \; t+a}^3  \right) .
\end{eqnarray}
Let us take now the function ${\cal G} (\alpha ) =
\frac{1}{2} ( \alpha - \sin \alpha )$,
being monotonically increasing as its first derivative
$ {\cal G} ' (\alpha ) = \sin^2 \frac{ \alpha }{2} >0$.  Then the
equation ${\cal G} (\beta ) = {\cal G} (\alpha )$ has only a single
solution $\beta = \alpha$. Making use of this property, we
 can write:
\begin{eqnarray}
\label{FtoG}
  \delta \left( {\cal F} ( aA_0^3 )_{t+a}  -
                {\cal F} ( aA_0^3 )_t        \right)
   &  {\rightarrow } &
   \frac{2 \sin^2 \left( \frac{ 1}{2} aA_{0 \; t} \right)  }{
           | {\cal F}' (aA_{0 \; t}^3  )  |  }
     \delta \left( {\cal G} ( aA_{0 \; t}^3 )
          - {\cal G} ( aA_{0 \; t+a}^3 )  \right) .
\end{eqnarray}
Inserting this into the physical vacuum to vacuum transition
amplitude, the terms $\sum_x \ln | {\cal F} '|$ cancel in the
exponent of the integrand, and we recover the Haar measure induced
tree level effective potential:
\begin{eqnarray}
   {\cal Z} \lbrack 0 \rbrack
   &=&
   \int {\cal D}_{\mbox {flat}} A_0^3 \int {\cal D} A_i^a
   \left( \prod_x
          \delta \left(    {\cal G} ( aA_0^3 )_{{\vec x} \; t+a}
               - {\cal G} ( aA_0^3 )_{ {\vec x}t}          \right)
   \right)
            \cdot        \nonumber\\
    & &  \cdot
      e^{  {\rm i} S \lbrack A^{ h_0 (A_0 ) \; j} , A^j  \rbrack
             + \sum_x \ln 2
             + \sum_x \ln \sin^2 \left( \frac{1}{2} aA_0^3 \right)
                            }  .
\end{eqnarray}

Shifting the argument of the Dirac delta
 by a (vector potential independent) constant   $c_x$
and integrate over $c_x$ by Gaussian weights, we obtain:
\begin{eqnarray}
\lefteqn{
   \int {\cal D} c \exp \left\{ - \frac{ a^2 }{ 2 g^2 \xi }
                    \sum_x c_x^2  \right\}  \ldots
  \left( \prod_x
      \delta \left(
        {\cal G} ( a A_{0 \; t+a}^3  )
                   - {\cal G} ( a A_{0 \; t}^3 )
        - c_x  \right)
              \right)
          =                 \qquad
          }  \nonumber\\
       &=&
   \exp \left\{ - \frac{ a^2}{ 2g^2 \xi } \sum_x \frac{1}{a^2}
    \left( {\cal G} (a A_{0 \; t+a}^3 )
         - {\cal G} ( a A_{0 \; t }^3 ) \right)^2
         \right\}   \ldots
\end{eqnarray}
Then  the physical vacuum to vacuum transition
amplitude takes the form:
\begin{eqnarray}
   {\cal Z} \lbrack 0 \rbrack
   &=&
   \int {\cal D}_{\mbox {flat}} A_0^3 \int {\cal D} A_i^a
    e^{  {\rm i} S \lbrack A^{ h_0 (A_0 ) \; j} , A^j  \rbrack
             + \sum_x \ln 2
             + \sum_x \ln \sin^2 \left( \frac{1}{2} aA_0^3 \right)
                            }       \cdot
               \nonumber\\
         & &   \cdot
   \exp \left\{ - \frac{ a^2}{ 2g^2 \xi } \sum_x \frac{1}{a^2}
    \left( {\cal G} (a A_{0 \; t+a}^3 )
         - {\cal G} ( a A_{0 \; t }^3 ) \right)^2
         \right\}   .
\end{eqnarray}
The additional interaction term expressed through the function
 ${\cal
G} (\alpha )$ is also defined completely by the Haar measure.

\section{Solution of the self-consistency and the saddle point
equations}
\setcounter{equation}{0}

We find the solution iteratively in two steps. At first we solve
equations (\ref{saddle}) and (\ref{selfc}) in the tree
 approximation.
Then we show that
the functional $h_{0x}$ and the values ${\bar u}$ are not modified
by the 1-loop corrections.

\begin{enumerate}

\item
In tree approximation the self-consistency equation (\ref{selfc})
\begin{eqnarray}
\label{sctree}
  \left.  \frac{ \delta \ln z \lbrack \xi \rbrack }{
          \delta \xi_x  }    \right|_{ \xi_x = h_{0x} \lbrack q=0
             \rbrack}            &=& 0
\end{eqnarray}
is consistent with the saddle point equation (\ref{saddle}),
\begin{eqnarray}
   \sum_y  a^2 D_{xy} h_{0y}  + \left. \frac{ \delta \ln z
   \lbrack \xi
                 \rbrack  }{ \delta \xi_x }
           \right|_{  \xi_x = h_{0x} \lbrack q \rbrack - q_x  }
                     &=& 0 ,
\end{eqnarray}
only if   the functional $h_{0x} \lbrack q \rbrack$ vanishes for
vanishing external source, i.e. $ h_{0x} \lbrack q=0 \rbrack \; =
\; 0$.  In light of that we expand $\ln z \lbrack \xi \rbrack$ in
Taylor series at $\xi_x \equiv 0$:
\begin{eqnarray}
\label{zTay}
  \ln z \lbrack \xi \rbrack &=& \ln z \lbrack 0 \rbrack +
   \sum_x \sum_{k=1}^\infty \frac{1}{k!} f_k \xi_x^k ,
\end{eqnarray}
where $f_k$ is the $k$-th functional derivative of $\ln z$
with respect to $\xi_x$ at $\xi_x \equiv 0$.  We take into account
the terms up to the order ${\cal O} ( \xi^5 )$.
It is easy to show that
the odd derivatives of $\ln I_0 $ vanish at $\xi_x \equiv 0$,
$\delta^{2k+1} ( \ln I_0 ) /\delta \xi_x^{2k+1}  |_{\xi \equiv 0}
=0$. From Eq. (\ref{sctree}) we get $f_1 =0$, which then means
\begin{eqnarray}
 \left. \frac{ \delta \ln z_0 \lbrack \xi \rbrack  }{ \delta \xi_x }
 \right|_{\xi =0} &= &    \left.
  \sum_{ \sigma \neq 0} v_\sigma
  \frac{  {\rm i} \sigma}{ ( \xi_x +  {\rm i} \sigma )^2 }
       e^{ {\rm i} \sigma ( a{\bar u} + \pi )  }
             \right|_{\xi =0}
                \nonumber\\
  &=&     \frac{1}{\rm i}
 \sum_{\sigma \neq 0} v_\sigma
       \frac{ \cos (a {\bar u} \sigma ) + {\rm i}
              \sin (a {\bar u} \sigma )   }{   \sigma }
        e^{ {\rm i} \sigma \pi  }
             = 0 .
\end{eqnarray}
Since $v_\sigma = v_{- \sigma}$, the real part of the sum is
automatically zero. The imaginary part,
\begin{eqnarray}
   \sum_{\sigma \neq 0} v_\sigma \frac{(- 1)^\sigma }{  \sigma}
   \sin ( a {\bar u} \sigma )  =0
\end{eqnarray}
vanishes for ${\bar u}$ satisfying $\sin (a {\bar u} \sigma ) =0$
for
any integer $\sigma$, i.e. for $ a {\bar u} = \nu \pi$, $(\nu =
{\mbox {integer}})$. Then we get $e^{ {\rm i}  a {\bar u} \sigma } =
(-1)^{\nu \sigma }$,  and
\begin{eqnarray}
   f_k &=& (-1)^{k+1} \sum_{\sigma \neq 0} v_\sigma (-1)^{\sigma
    (\nu
    +1) }    \frac{ k!}{ ( {\rm i} \sigma )^k } ,
\end{eqnarray}
i.e.  all the odd derivatives of $\ln z$
vanish at $\xi =0$, $f_{2k+1} =0$, and the even derivatives are
given by:
\begin{eqnarray}
  f_{2k} &=& (-1)^{k+1} 2 \sum_{\sigma >0} v_\sigma
    (-1)^{\sigma ( \nu +1) }
        \frac{ (2k)!}{  \sigma ^{2k} } .
\end{eqnarray}
Note that $\nu = \pm 1$ correspond to the minima of the
effective potential in the tree approximation. For $\nu = \pm 1$
holds:
\begin{eqnarray}
   f_{2k} &=& (-1)^{k+1} \cdot 2 \sum_{\sigma > 0 } v_\sigma \frac{
 (2k)!  }{  \sigma^{2k} } ,
\end{eqnarray}
i.e. $f_2 >0$, $f_4 < 0$, etc. We shall see that these are the
 correct
signs ensuring that the effective potential at 1--loop order
 has minima at
$a{\bar u} = \pm \pi$.

We rewrite the saddle point equation (\ref{saddle}):
\begin{eqnarray}
  \sum_y \lbrack  a^2 D_{xy} + f_2 \delta_{xy} \rbrack \xi_y
    &=& - \sum_y a^2 D_{xy} q_y - \frac{1}{6} f_4 \xi_x^3
\end{eqnarray}
including terms up to the order ${\cal O} (\xi^4 )$.
Introducing the matrix
\begin{eqnarray}
  (a^2 Q_0 )_{xy}  &=&  a^2 D_{xy} + f_2 \delta_{xy} ,
\end{eqnarray}
we can write:
\begin{eqnarray}
  \xi_x &= & - \sum_{yz} \lbrack (a^2 Q_0 )^{-1} \rbrack_{xy}
          a^2 D_{yz} q_z
   - \frac{1}{6} f_4 \sum_z \lbrack (a^2 Q_0 )^{-1} \rbrack_{xz}
             \xi_z^3  .
\end{eqnarray}
We solve this equation by iteration:
\begin{eqnarray}
  \xi_x & =& - \sum_{yz} \lbrack (a^2 Q_0 )^{-1} \rbrack_{xy}
          a^2 D_{yz} q_z
            \nonumber\\
  & & + \frac{1}{6} f_4 \sum_y \lbrack (a^2 Q_0 )^{-1} \rbrack_{xy}
             \left( \sum_{zu} \lbrack (a^2 Q_0 )^{-1} \rbrack_{yz}
             a^2 D_{zu} q_u  \right)^3
              \nonumber\\
  & & + {\cal O} (q^5 ) .
\end{eqnarray}
(Note that the matrices $D_{xy}$ and $(Q_0 )_{xy}$ are symmetric.)
For the sake of simplicity we introduce the notations
$ (Aq)_x = \sum_y A_{xy} q_y $,
$ (qAq) = \sum_{xy} q_x A_{xy} q_y$ and do not write out
 the powers of $a$
(they can be introduced at any step to obtain dimensionless
expressions),  then:
\begin{eqnarray}
\label{solsp}
   \xi_x &=& - (Q_0^{-1} D q )_x
    + \frac{1}{6} f_4  ( Q_0^{-1} (Q_0^{-1} Dq)^3 )_x  .
\end{eqnarray}

Including terms up to the order ${\cal O} (q^4 )$, we obtain:
\begin{eqnarray}
  \xi_x^2  &=& ( Q_0^{-1} D q)_x^2 -
    \frac{1}{3} f_4  (Q_0^{-1} Dq )_x  (Q_0^{-1} (
     Q_0^{-1} Dq)^3 )_x ,       \nonumber\\
   \xi_x^3 &=& - (Q_0^{-1} Dq )_x^3 ,
             \nonumber\\
   \xi_x^4 &=& (Q_0^{-1} Dq )_x^4
\end{eqnarray}
for later use.

\item

Now we show that the functional $h_{0x} \lbrack q \rbrack$ is not
 modified by 1-loop corrections. Including the 1-loop term the
 self-consistency  equation (\ref{selfc}) is given by:
\begin{eqnarray}
 f_2 \xi_x \lbrack 0 \rbrack + \frac{1}{6} f_4 \xi_x^3 \lbrack 0
   \rbrack  + \left. \frac{1}{2} \frac{ \delta }{\delta q_x }
   {\mbox {Tr}} \ln (a^2 Q )  \right|_{ q=0}  &=& 0 .
\end{eqnarray}
Inserting the explicit form of the 1-loop contribution of the
generating functional (Appendix E),
\begin{eqnarray}
W_1 \lbrack q \rbrack &= &
  - \frac{1}{2} {\mbox {Tr}} \ln (a^2 Q)  =
 \frac{f_4}{ 2 f_2} \mu_2 \sum_y ( Q_0^{-1} D q )^2_y  ,
\end{eqnarray}
and writing $ \xi_x \lbrack q \rbrack = \xi^{(0)} \lbrack q \rbrack
  + \hbar \xi^{(1)} \lbrack q \rbrack $, we obtain for the term
of order $\hbar$:
\begin{eqnarray}
 \left.  \xi_x^{(1)} \lbrack q \rbrack  \right|_{q=0}
   &=& - \left. \frac{1}{2f_2} \frac{ \delta}{\delta q_x}
    {\mbox {Tr}}
    \ln (a^2 Q ) \right|_{q=0}
   =  \left.
 \frac{1}{f_2} \frac{\delta}{ \delta q_x} W_1 \lbrack q \rbrack
    \right|_{q=0}
             \nonumber\\
  &=& \frac{f_4}{f_2^2} \mu_2 \left.  \left(  (Q_0^{-1} D)^2 q
                 \right)_x   \right|_{q=0}  = 0.
\end{eqnarray}
This means that the initial value $h_{0x} \lbrack q=0 \rbrack =0$
 is not
modified due to 1-loop corrections.  Then the solution of the saddle
point equation remains  unaltered and also the value of $a {\bar u}$
is not modified.

\end{enumerate}

\section{1--loop contribution }
\setcounter{equation}{0}

Now  we calculate the 1--loop contribution $W_1 \lbrack q \rbrack$
to the generating  functional of the skelet model.

The 1--loop part of the generating functional of the connected
 Green's
functions  can be rewritten:
\begin{eqnarray}
  W_1 \lbrack q \rbrack &=&  {\mbox {const.}}
       - \frac{1}{2} {\mbox {Tr}}
     \ln ( a^2 Q ) = {\mbox {const.}}
      - \frac{1}{2} {\mbox {Tr}} \ln \left\lbrack  (a^2 D)
      \left( 1 + f_2 (\xi )
         D^{-1}  \right)  \right\rbrack
                   \nonumber\\
    & = & {\mbox {const.}} - \frac{1}{2} {\mbox {Tr}}
        \ln \left(  1 + f_2 (\xi )  D^{-1}  \right)
                    \nonumber\\
    &=& {\mbox {const.}}  + \frac{1}{2}
         \sum_{ n=1}^\infty \frac{ (-1)^n }{n} {\mbox {Tr}}
         \left\lbrack f_2 (\xi ) D^{-1}  \right\rbrack^n ,
\end{eqnarray}
where $f_2 (\xi )$ is defined by
 $ \frac{ \delta^2}{ \delta \xi_x \delta \xi_y } \ln z \lbrack
\xi \rbrack = \delta_{xy} f_2 (\xi_x ) $.  Making use of the Taylor
expansion (\ref{zTay}) with vanishing odd derivatives, we get for
 the
trace:
\begin{eqnarray}
  {\mbox { Tr}} \lbrack f_2 (\xi ) D^{-1} \rbrack^n
    &=& {\mbox {Tr}} \left\{ f_2^n ( D^{-1} )^n
   + n f_2^{n-1} ( D^{-1} )^{n-1}
     \left\lbrack \frac{1}{2} f_4 ( \xi^2 D^{-1} ) +
      \frac{1}{24} f_6 ( \xi^4 D^{-1} )  \right\rbrack
             \right.
               \nonumber\\
   & & \left. + \frac{1}{2} n (n-1) f_2^{n-2} ( D^{-1} )^{n-2}
                \frac{1}{4} f_4^2 ( \xi^2 D^{-1} )^2
                + {\cal O} ( \xi^6 )
        \right\}  ,
\end{eqnarray}
and for $W_1 \lbrack q \rbrack$:
\begin{eqnarray}
  W_1 \lbrack q \rbrack
    &=&  {\mbox {const.}}
  + \frac{1}{4} \frac{f_4}{f_2} \sum_{n=1}^\infty (-1)^n f_2^{n}
    \sum_x \xi_x^2 \left\lbrack  (D^{-1} )^{n} \right\rbrack_{xx}
                     \nonumber\\
   & & + \frac{1}{48} \frac{f_6}{f_2} \sum_{n=1}^\infty
    (-1)^n f_2^{n}
    \sum_x \xi_x^4 \left\lbrack (D^{-1} )^{n} \right\rbrack_{xx}
                      \nonumber\\
    & & + \frac{1}{16} f_4^2 \sum_{n=1}^\infty (-1)^n (n-1)
        f_2^{n-2} \sum_{xyz} \xi_x^2 (D^{-1} )_{xy}
                             \xi_z^2 (D^{-1} )_{zy}
        \left\lbrack (D^{-1} )^{n-2}  \right\rbrack_{yy} .
                     \nonumber\\
\end{eqnarray}
Let us make use of the fact that the diagonal matrix elements of the
powers of the  inverse propagator do not depend on the Euclidean
coordinates due to translation invariance (in the infinite volume
limit) and reexpress the sums over $n$ in terms of $\mu_2$:
\begin{eqnarray}
   \mu_2 &=& \frac{1}{2} \sum_{n=1}^\infty (-1 )^n f_2^n \lbrack
   ( D^{-1}  )^n \rbrack_{xx} .
\end{eqnarray}
The last term on the r.h.s. of $W_1$ we rewrite as follows:
\begin{eqnarray}
\lefteqn{
  \frac{1}{16} f_4^2
    \left( \xi^2 (D^{-1} )^2 \xi^2 \right)
     \sum_{n=1}^\infty (-1)^n (n-1) f_2^{n-2}
     \lbrack ( D^{-1} )^{n-2} \rbrack_{uu}
             }        \nonumber\\
 &= & \frac{1}{16} f_4^2
    \left( \xi^2 (D^{-1} )^2 \xi^2 \right)
   (-1) \sum_{m=0}^\infty (-1)^m m f_2^{m-1} \lbrack (D^{-1} )^{m-1}
         \rbrack_{uu}
                      \nonumber\\
 &= & \frac{1}{16} f_4^2
    \left( \xi^2 (D^{-1} )^2 \xi^2 \right)
   (-1) \sum_{m=1}^\infty (-1)^m m f_2^{m-1} \lbrack (D^{-1} )^{m-1}
         \rbrack_{uu}
                      \nonumber\\
 &= & \frac{1}{16} f_4^2
    \left( \xi^2 (D^{-1} )^2 \xi^2 \right)
   (-1) \frac{ \partial }{\partial f_2} \left\lbrack - f_2
 \sum_{m=1}^\infty (-1)^{m-1} f_2^{m-1} \lbrack (D^{-1} )^{m-1}
         \rbrack_{uu}                   \right\rbrack
                       \nonumber\\
 &= & \frac{1}{16} f_4^2
    \left( \xi^2 (D^{-1} )^2 \xi^2 \right)
    \frac{ \partial }{\partial f_2} \left\lbrack  f_2
 \sum_{n=0}^\infty (-1)^{n} f_2^{n} \lbrack (D^{-1} )^{n}
         \rbrack_{uu}                   \right\rbrack
                       \nonumber\\
 &= & \frac{1}{16} f_4^2
    \left( \xi^2 (D^{-1} )^2 \xi^2 \right)
    \frac{ \partial }{\partial f_2} \left\lbrack  f_2
 \sum_{n=1}^\infty (-1)^{n} f_2^{n} \lbrack (D^{-1} )^{n}
         \rbrack_{uu}      + f_2             \right\rbrack
                       \nonumber\\
  &=& \frac{1}{8} f_4^2 \left( \frac{1}{2} +
   \frac{\partial}{\partial
       f_2}  (f_2 \mu_2 )  \right)  \left( \xi^2 (D^{-1} )^2 \xi^2
         \right) .
\end{eqnarray}
Here we can write:
\begin{eqnarray}
   \frac{\partial }{\partial f_2 } ( f_2 \mu_2 )
   &=& - \frac{ \partial}{\partial f_2} \left( \frac{ f_2^2}{g^2}
          \right)  = - \frac{2f_2}{g^2} = 2\mu_2 .
\end{eqnarray}
Finally the 1--loop contribution to the generating functional of the
connected Green's functions takes the form:
\begin{eqnarray}
  W_1 \lbrack \xi \rbrack &=& {\mbox {const.}}
    + \frac{1}{2} \frac{f_4}{f_2} \mu_2 \sum_x \xi_x^2
    + \frac{1}{24} \frac{f_6}{f_2} \mu_2 \sum_x \xi_x^4
                    \nonumber\\
   & & + \frac{1}{8} f_4^2 \left( \frac{1}{2} + 2\mu_2  \right)
     \sum_{xy} \xi^2_x \lbrack ( D^{-1} )^2 \rbrack_{xy} \xi^2_y .
\end{eqnarray}
Let us express this functional through the external source $q_x$:
\begin{eqnarray}
  W_1 \lbrack q \rbrack &=&
   \frac{1}{2} \frac{f_4}{f_2} \mu_2 \sum_x \left\lbrack
   ( Q_0^{-1} D q )_x^2 - \frac{1}{3} f_4 ( Q_0^{-1} D q )_x
   \left( Q_0^{-1} ( Q_0^{-1} D q)^3  \right)_x   \right\rbrack
             \nonumber\\
   & & + \frac{1}{24} \frac{f_6}{f_2} \mu_2
    \sum_x ( Q_0^{-1} Dq)_x^4
            \nonumber\\
   & & + \frac{1}{8} g_4^2 \sum_{xy} (Q_0^{-1} Dq)_x^2
     (D^{-1} )^2_{xy}  (Q_0^{-1} Dq )^2_y ,
\end{eqnarray}
where  $g_4^2 = f_4^2 \left( \frac{1}{2} + 2 \mu_2 \right)$.

Now we see that the 1--loop contribution is determined by
$\mu_2$:
\begin{eqnarray}
  \mu_2 &=& \frac{1}{2}  \sum_{n=1}^\infty (-1)^n f_2^n a^4
    \int_B \frac{d^4 p}{ (2\pi )^4 } \left\lbrack a^2
     {\tilde D}^{-1}         (p)  \right\rbrack^n
                   \nonumber\\
      &=& \frac{1}{2} \left\{ \sum_{n=0}^\infty (-1)^n
    \left( \frac{2f_2}{g^2} \right)^n a^4 \int_B
    \frac{d^4 p}{ (2\pi^4 )} \left\lbrack \sum_{\mu =0}^3
     \left( 1 - \cos (ap^\mu ) \right) \right\rbrack^n
          - 1 \right\}
                    \nonumber\\
      &=& \frac{1}{2} \left\{ a^4 \int_B \frac{d^4 p}{ (2\pi )^4}
   \frac{1}{  1 + 2f_2 g^{-2} \sum_{\mu =0}^3 \left( 1 - \cos (
             ap^\mu )     \right)   }
        - 1           \right\} .
\end{eqnarray}
The geometric series converges  for $8s \equiv 8 \cdot 2f_2 g^{-2}
  \le 1$, i.e. $s \le \frac{1}{8}$.
We write the nominator of the integrand as:
$4s \left( 1 - {\tilde \beta} \sum_{\mu =0}^3 \cos (ap^\mu )
 \right)$
with ${\tilde \beta} = s (1+s )^{-1}$. It is well
 known\cite{Par??} that the
momentum integral converges for $0 < {\tilde \beta} <
\frac{1}{4}$ which is satisfied for $s \le \frac{1}{8}$.

Let us perform the integral over $p^0$ at first. The integrand
develops simple poles on the complex $p^0$ plane if ${\vec p} \to 0$
\cite{Par??}.
We estimate the momentum integral by  the pole contribution, and set
${\vec p} =0$. Then the 3-momentum integration can be carried out
trivially, and the 1--dimensional integral
\begin{eqnarray}
   \mu_2 &=& \frac{1}{2} \left(  - \frac{ 1}{(1+4s) {\tilde \beta} }
    \frac{a}{2\pi}  \int_{-\pi /a}^{\pi /a} dp^0
    \frac{ 1 }{ \zeta -   \cos (ap^0 ) }    - 1   \right)
\end{eqnarray}
is left over, $\zeta = ( 1 - 3 {\tilde \beta} ) {\tilde \beta}^{-1}
 = s^{-1} +1$.
It has simple poles at $ap^0 = {\rm i} v_\pm = {\rm i}
  \ln \left( \zeta \pm \sqrt{  \zeta^2 - 1 }  \right)$.
We can take $v_\pm \approx \pm \ln ( 2\zeta )$ considering
 $\zeta \gg
1$ for $s \le \frac{1}{8}$. The 1-dimensional integral is
of the same
type as the integral ${\cal I}_0$ in Part V. With a similar
treatment,
 we obtain:
\begin{eqnarray}
   \mu_2 &=& \frac{1}{2} \left(   \frac{ 1}{s \sinh v_+ } -1
          \right)
 = \frac{1}{2} \left(   \frac{ 1}{s \zeta } -1
          \right)
     \approx - \frac{1}{2} s  .
\end{eqnarray}
The parameter $s \ll 1$ is a small one in the theory.

The generating functional $W \lbrack q \rbrack$ of the connected
 Green's functions is the sum of the tree contribution
  $W_0 \lbrack q
\rbrack = \ln {\cal Z}_0 + {\mbox {const.}}$ and that of the 1-loop
contribution $W_1 \lbrack q \rbrack$:
\begin{eqnarray}
\label{Wq}
  W \lbrack q \rbrack &=&
    - a{\bar u} \sum_x q_x + \frac{1}{2} f_2 ( q Q_1 Q_0^{-1} D q )
       \nonumber\\
    & & - \frac{1}{24} f_4 \left( (Q_0^{-1} D q )^4 \right)
     + \frac{1}{12} f_4 \left( q ( 1 -f_2 Q_0^{-1} Q_2 )
         ( Q_0^{-1} D q)^3   \right)
             \nonumber\\
     & & + \frac{1}{16} f_4^2 \sum_{xy} (Q_0^{-1} D q)_x^2
    ( D^{-1} )_{xy}^2 ( Q_0^{-1} Dq )_y^2 ,
\end{eqnarray}
with
\begin{eqnarray}
   Q_0 = D + f_2 = D ( 1 + f_2 D^{-1} ) ,
                \nonumber\\
   Q_1 = 1 + \lambda Q_0^{-1} D ,  \qquad
   Q_2 = 1 + 2 \lambda Q_0^{-1} D  ,
\end{eqnarray}
and the parameter
\begin{eqnarray}
   \lambda &=& \frac{f_4}{f_2^2 } \mu_2 = - \frac{f_4 }{ 2f_2^2 } s
     = - \frac{f_4}{f_2 g^2}  .
\end{eqnarray}
The parameter $\lambda$ governs the loop expansion of the generating
functional $W \lbrack q \rbrack$. The terms
proportional to $\lambda$ in $Q_1$ and $Q_2$ represent the 1-loop
contributions.  The last term on the r.h.s. of Eq. (\ref{Wq})
is of the order $f_4^2 f_2^{-2} s^2 \sim f_2^2 \lambda^2$ and
was neglected.
 (We neglected also the term proportional to $f_6$.)

\end{document}